# SETI, Evolution and Human History Merged into a Mathematical Model

Claudio Maccone

*International Academy of Astronautics (IAA), Via Martorelli, 43, Torino (Turin) 10155, Italy*
*e-mail: clmaccon@libero.it and claudio.maccone@iaamail.org*

**Abstract**: In this paper we propose a new mathematical model capable of merging Darwinian Evolution, Human History and SETI into a single mathematical scheme:
(1) Darwinian Evolution over the last 3.5 billion years is defined as one particular realization of a certain stochastic process called Geometric Brownian Motion (GBM). This GBM yields the fluctuations in time of the number of species living on Earth. Its mean value curve is an increasing exponential curve, i.e. the exponential growth of Evolution.
(2) In 2008 this author provided the statistical generalization of the Drake equation yielding the number *N* of communicating ET civilizations in the Galaxy. *N* was shown to follow the lognormal probability distribution.
(3) We call "*b*-lognormals" those lognormals starting at any positive time *b* ("birth") larger than zero. Then the exponential growth curve becomes the geometric locus of the peaks of a one-parameter family of *b*-lognormals: this is our way to re-define Cladistics.
(4) *b*-lognormals may be also be interpreted as the lifespan of any living being (a cell, or an animal, a plant, a human, or even the historic lifetime of any civilization). Applying this new mathematical apparatus to Human History, leads to the discovery of the exponential progress between Ancient Greece and the current USA as the envelope of all *b*-lognormals of Western Civilizations over a period of 2500 years.
(5) We then invoke Shannon's Information Theory. The *b*-lognormals' entropy turns out to be the index of "development level" reached by each historic civilization. We thus get a numerical estimate of the entropy difference between any two civilizations, like the Aztec-Spaniard difference in 1519.
(6) In conclusion, we have derived a mathematical scheme capable of estimating how much more advanced than Humans an Alien Civilization will be when the SETI scientists will detect the first hints about ETs.

## SETI and Darwinian Evolution Merged Mathematically

*Introduction: the Drake equation (1961) as the Foundation of SETI*

In 1961, the American astronomer Frank D. Drake tried to estimate the number *N* of communicating civilizations in the Milky Way galaxy by virtue of a simple equation now called the Drake equation. *N* was written as the product of seven factors, each a kind of filter, every one of which must be sizable for there to be a large number of civilizations: $Ns$, the number of stars in the Milky Way Galaxy; $fp$, the fraction of stars that have planetary systems; $ne$, the number of planets in a given system that are ecologically suitable for life; $fl$, the fraction of otherwise suitable planets on which life actually arises; $fi$, the fraction of inhabited planets on which an intelligent form of life evolves (as in Human History); $fc$, the fraction of planets inhabited by intelligent beings on which a communicative technical civilization develops (as we have it today); and $fL$, the fraction of planetary lifetime graced by a technical civilization (a totally unknown factor).

Written out, the equation reads

$$N = Ns \cdot fp \cdot ne \cdot fl \cdot fi \cdot fc \cdot fL. \qquad (1)$$

All the *f*'s are fractions, having values between 0 and 1; they will pare down the large value of *Ns*. To derive *N*, we must estimate each of these quantities. We know a fair amount about the early factors in the equation, the number of stars and planetary systems. We know very little about the later factors, concerning the evolution of life, the evolution of intelligence or the lifetime of technical societies. In these cases, our estimates will be little better than guesses.

It has to be said that the original formulation of (1) by Frank Drake in 1961 was slightly different, namely

$$N = R^* \cdot fp \cdot ne \cdot fl \cdot fi \cdot fc \cdot L. \qquad (2)$$

In (2), $R^*$ is the average rate of star formation per year in the Galaxy and *L* is the length of time for which civilizations in the Galaxy release detectable signals into space. However, the number of stars in the Galaxy, *Ns*, is related to the star formation rate $R^*$ by

$$Ns = \int_0^{T_{\text{Galaxy}}} R^*(t)\,dt, \tag{3}$$

where $T_{\text{Galaxy}}$ is the age of the Galaxy. Assuming for simplicity that $R^*$ is constant in time, then (3) yields

$$Ns = R^* \cdot T_{\text{Galaxy}} \quad \text{i.e.} \quad R^* = \frac{Ns}{T_{\text{Galaxy}}}, \tag{4}$$

that, inserted into (2), changes it into

$$N = Ns \cdot fp \cdot ne \cdot fl \cdot fi \cdot fc \cdot \frac{L}{T_{\text{Galaxy}}}. \tag{5}$$

Then (5) becomes just (1) if one identifies

$$fL = \frac{L}{T_{\text{Galaxy}}} \tag{6}$$

as the fraction of planetary lifetime (as a part of the whole Galaxy existence $T_{\text{Galaxy}}$) graced by a technical civilization.

In the 50 years that have elapsed since Drake proposed his equation, a number of scientists and writers have tried either to improve it or criticize it in many ways. For instance, in 1980, C. Walters, R. A. Hoover and R. K. Kotra (Walters *et al.*, 1980) suggested inserting a new parameter in the equation taking interstellar colonization into account. In 1981, S. G. Wallenhorst (Wallenhorst 1981) tried to prove that there should be an upper limit of about 100 to the number *N*. In 2004, L. V. Ksanfomality (Ksanfomality 2004) again asked for more new factors to be inserted into the Drake equation, this time in order to make it compatible with the peculiarities of planets of Sun-like stars. Also, the temporal aspect of the Drake equation was stressed by Ćirković (2004). However, while these authors were concerned with improving the Drake equation, others simply did not consider it useful and preferred to forget about it, like Burchell (2006).

Also, it has been correctly pointed out that the habitable part of the Galaxy is probably much smaller than the entire volume of the Galaxy itself (the important relevant references are Gonzalez *et al.* (2001), Lineweaver *et al.* (2004) and Gonzalez (2005)). For instance, it might be a sort of torus centred around the so-called 'corotation circle', i.e. a circle around the Galactic Bulge such that stars orbiting around the Bulge and within such a torus never fall inside the dangerous spiral arms of the Galaxy, where supernova explosions would probably fry any living organism before it could develop to the human level or beyond. Fortunately for Humans, the orbit of the Sun around the Bulge is just a circle staying within this torus for 5 billion years or more (Marochnik & Mukhin 1988; Balazs 1988).

In all cases, the final result about *N* has always been a sheer number, i.e. a positive integer number ranging from 1 to thousands or millions. This 'integer or real number' aspect of all variables making up the Drake equation is what this author regarded as 'too simplistic'. He extended the Drake equation so as to embrace Statistics in his 2008 paper (Maccone 2008).

This paper was later published in *Acta Astronautica* (Maccone 2010a), and more mathematical consequences were derived in Maccone (2010b) and Maccone (2011a).

*Statistical Drake equation (2008)*

Consider *Ns*, the number of stars in the Milky Way Galaxy, i.e. the first independent variable in the Drake equation (1). Astronomers tell us that approximately there should be about 350 billion stars in the Galaxy. Of course, nobody has counted all the stars in the Galaxy! There are too many practical difficulties preventing us from doing so: just to name one, the dust clouds that do not allow us to see even the Galactic Bulge (central region of the Galaxy) in visible light, although we may 'see it' at radio frequencies like the famous neutral hydrogen line at 1420 MHz. Hence, it does not really make much sense to say that $Ns = 350 \times 10^9$, or similar fanciful exact integer numbers. Scientifically we say that the number of stars in the Galaxy is 350 billion plus or minus, say, 50 billions (or whatever values the astronomers may regard as more appropriate).

It thus makes sense to REPLACE each of the seven independent variables in the Drake equation (1) by a mean value (350 billions, in the above example) plus or minus a certain standard deviation (50 billions, in the above example).

By doing so, we moved a step ahead: we have abandoned the too-simplistic equation (1) and replaced it by something more sophisticated and scientifically serious: the statistical Drake equation. In other words, we have transformed the simplistic classical Drake equation (1) into a statistical tool capable of investigating a host of facts hardly known to us in detail. In other words still:

(1) we replace each independent variable in (1) by a random variable, labelled $D_i$ (from Drake);
(2) we assume the mean value of each $D_i$ to be the same numerical value previously attributed to the corresponding input variable in (1);
(3) but now we also add a standard deviation $\sigma_{D_i}$ on each side of this mean value, as provided by the knowledge obtained by scientists in the discipline covered by each $D_i$.

Having done so, we wonder: how can we find out the probability distribution for each $D_i$? For instance, shall that be a Gaussian, or what? This is a difficult question, for nobody knows, for instance, the probability distribution of the number of stars in the Galaxy, not to mention the probability distribution of the other six variables in the Drake equation (1). In 2008, however, this author found a way to get around this difficulty, as explained in the next section.

*The statistical distribution of N is lognormal*

The solution to the problem of finding the analytical expression for the probability density function (pdf) of the positive random variable *N* is as follows:

(1) Take the natural logs of both sides of the statistical Drake equation (1). This changes the product into a sum.
(2) The mean values and standard deviations of the logs of the random variables $D_i$ may all be expressed analytically in terms of the mean values and standard deviations of $D_i$ (Maccone 2008).

Table 1. *Summary of the properties of the lognormal distribution that applies to the random variable N = number of ET communicating civilizations in the Galaxy*

| Random variable | $N$ = number of communicating ET civilizations in Galaxy |
|---|---|
| Probability distribution | lognormal |
| Pdf | $f_N(n) = \frac{1}{n} \frac{1}{\sqrt{2\pi}\sigma} e^{-\frac{(\ln(n)-\mu)^2}{2\sigma^2}} \quad (n \geq 0)$ |
| Mean value | $\langle N \rangle = e^\mu e^{\frac{\sigma^2}{2}}$ |
| Variance | $\sigma_N^2 = e^{2\mu} e^{\sigma^2}(e^{\sigma^2} - 1)$ |
| Standard deviation | $\sigma_N = e^\mu e^{\frac{\sigma^2}{2}} \sqrt{e^{\sigma^2} - 1}$ |
| All the moments, i.e. $k$th moment | $\langle N^k \rangle = e^{k\mu} e^{k^2 \cdot \frac{\sigma^2}{2}}$ |
| Mode (= abscissa of the lognormal peak) | $n_{\text{mode}} \equiv n_{\text{peak}} = e^\mu e^{-\sigma^2}$ |
| Value of the mode peak | $f_N(n_{\text{mode}}) = \frac{1}{\sqrt{2\pi}\sigma} e^{-\mu} e^{\frac{\sigma^2}{2}}$ |
| Median (= fifty–fifty probability value for $N$) | Median = $m = e^\mu$ |
| Skewness | $\frac{K_3}{(K_2)^{\frac{3}{2}}} = (e^{\sigma^2} + 2)\sqrt{e^{\sigma^2} - 1}$ |
| Kurtosis | $\frac{K_4}{(K_2)^2} = e^{4\sigma^2} + 2e^{3\sigma^2} + 3e^{2\sigma^2} - 6$ |
| Expression of $\mu$ in terms of the lower ($a_i$) and upper ($b_i$) limits of the Drake uniform input random variable $D_i$ | $\mu = \sum_{i=1}^{7} \langle Y_i \rangle = \sum_{i=1}^{7} \frac{b_i[\ln(b_i) - 1] - a_i[\ln(a_i) - 1]}{b_i - a_i}$ |
| Expression of $\sigma^2$ in terms of the lower ($a_i$) and upper ($b_i$) limits of the Drake uniform input random variable $D_i$ | $\sigma^2 = \sum_{i=1}^{7} \sigma_{Y_i}^2 = \sum_{i=1}^{7} \left(1 - \frac{a_i b_i [\ln(b_i) - \ln(a_i)]^2}{(b_i - a_i)^2}\right)$ |

(3) The central limit theorem (CLT) of statistics, states that (loosely speaking) if you have a sum of independent random variables, each of which is arbitrarily distributed (hence, also including uniformly distributed), then, when the number of terms in the sum increases indefinitely (i.e. for a sum of random variables infinitely long)... the sum random variable approaches a Gaussian.
(4) Thus, the ln(N) approaches a Gaussian.
(5) Namely, $N$ approaches the lognormal distribution (as discovered back in the 1870s by Sir Francis Galton). Table 1 shows the most important statistical properties of a lognormal.
(6) The mean value and standard deviations of this lognormal distribution of $N$ may be expressed analytically in terms of the mean values and standard deviations of the logs of $D_i$ already found previously, as shown in Table 1.

For all the relevant mathematical proofs, more mathematical details and a few numerical examples of how the Statistical Drake Equation works, please see Maccone (2010a).

*Darwinian evolution as exponential increase of the number of living species*

Consider now Darwinian Evolution. To assume that the number of species increased exponentially over 3.5 billion years of evolutionary time span is certainly a gross oversimplification of the real situation, as proven, for instance, by Rohde & Muller (2005). However, we will assume this exponential increase of the number of living species in time just for a moment in order to cast the theory into a mathematically simple and fruitful form. The introduction of Geometric Brownian Motion (GBM) in the next section of this paper will solve this difficulty.

In other words, we assume that 3.5 billion years ago there was on Earth only one living species, whereas now there may be (say) 50 million living species or more. Note that the actual number of species currently living on earth does not really matter as a number for us: we just want to stress the exponential character of the growth of species. Thus, we shall assume that the number of living species on Earth increases in time as $E(t)$ (standing for 'exponential in time'):

$$E(t) = Ae^{Bt}, \qquad (7)$$

where $A$ and $B$ are two positive constants that we will soon determine numerically. This assumption of ours is obviously in agreement with the classical Malthusian theory of population growth. However, it also is in line with the more recent 'Big History' viewpoint about the whole evolution of the Universe, from the Big Bang up to now, requesting that progress in evolution occurs faster and faster, so that only an exponential growth is compatible with the requirements that (7) approaches infinity for $t \to \infty$ and all its time derivatives are exponentials too, apart from constant multiplicative factors.

Let us now adopt the convention that the current epoch corresponds to the origin of the time axis, i.e. to the instant $t = 0$. This means that all the past epochs of Darwinian Evolution correspond to negative times, whereas the future ahead of us (including finding ETs) corresponds to positive times. Setting $t = 0$ in (7), we immediately find

$$E(0) = A \qquad (8)$$

proving that the constant $A$ equals the number of living species on earth right now. We shall assume

$$A = 50 \text{ million species} = 5 \times 10^7 \text{ species}. \qquad (9)$$

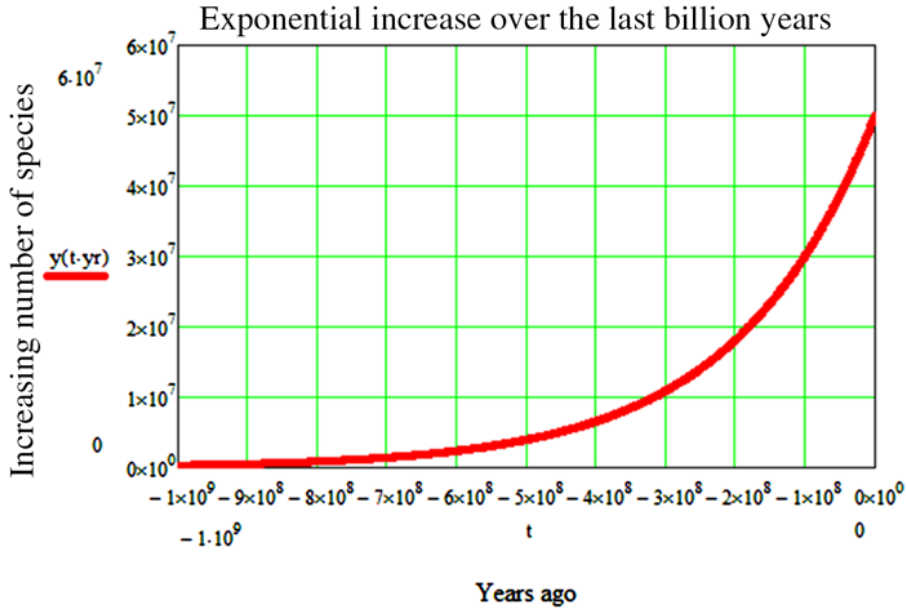

**Fig. 1.** Exponential curve representing the growing number of species on Earth up to now, without taking the well-known Mass Extinctions into any consideration at all.

To also determine the constant $B$ numerically, consider two values of the exponential (7) at two different instants $t_1$ and $t_2$, with $t_1 < t_2$, that is

$$\begin{cases} E(t_1) = Ae^{Bt_1}, \\ E(t_2) = Ae^{Bt_2}. \end{cases} \quad (10)$$

Dividing the lower equation by the upper one, $A$ disappears and we are left with an equation in $B$ only:

$$\frac{E(t_2)}{E(t_1)} = e^{B(t_2 - t_1)}. \quad (11)$$

Solving this for $B$ yields

$$B = \frac{\ln(E(t_2)) - \ln(E(t_1))}{t_2 - t_1}. \quad (12)$$

We may now impose the initial condition stating that 3.5 billion years ago there was just one species on Earth, the first one (whether this was RNA is unimportant in the present simple mathematical formulation):

$$\begin{cases} t_1 = -3.5 \times 10^9 \text{ years}, \\ E(t_1) = 1 \quad \text{whence} \quad \ln(E(t_1)) = \ln(1) = 0. \end{cases} \quad (13)$$

The final condition is of course that today ($t_2 = 0$) the number of species equals $A$ given by (9). Upon replacing both (9) and (13) into (12), the latter becomes:

$$B = -\frac{\ln(E(t_2))}{t_1} = -\frac{\ln(5 \times 10^7)}{-3.5 \times 10^9 \text{ year}} = \frac{1.605 \times 10^{-16}}{\text{sec}}. \quad (14)$$

Having thus determined the numerical values of both $A$ and $B$, the exponential in (7) is thus fully specified. This curve is plotted in Figure 1 just over the last billion years, rather than over the full range between $-3.5$ billion years and now.

*Introducing the 'Darwin' (d) unit, measuring the amount of evolution that a given species reached*

In all sciences 'to measure is to understand'.

In physics and chemistry this is done by virtue of units such as the metre, second, kilogram, coulomb, etc. Hence, it appears useful to introduce a new unit measuring the degree of evolution that a certain species has reached at a certain time $t$ of Darwinian Evolution, and the obvious name for such a new unit is the 'Darwin', denoted by a lower case 'd'. For instance, if we adopt the exponential evolution curve described in the previous section, we might say that the dominant species on Earth right now (Humans) have reached an evolution level of 50 million Darwins.

How many Darwins may have an alien civilization already reached? Certainly more than 50 millions, i.e. more than 50 Md, but we will not check out until SETI scientists will possibly detect the first extraterrestrial civilization.

We are not going to discuss further this notion of measuring the 'amount of evolution', since we are aware that endless discussions might come out of it. However, it is clear to us that such a new measuring unit (and ways to measure it for different species) will sooner or later have to be introduced to make Evolution a fully quantitative science.

*Darwinian evolution is just a particular realization of geometric Brownian motion in the number of living species*

Consider again the exponential curve described in the previous section. The most frequent question that non-mathematically minded persons ask this author is: 'then you do not take the mass extinctions into account'. The answer to this objection is that our exponential curve is just the mean value of a certain stochastic process that may run above and below that exponential in a totally unpredictable way. Such a stochastic process is called Geometric Brownian Motion (abbreviated GBM) and is described, for instance, at the web site: http://en.wikipedia.org/wiki/Geometric_Brownian_motion, from which Figure 2 is taken.

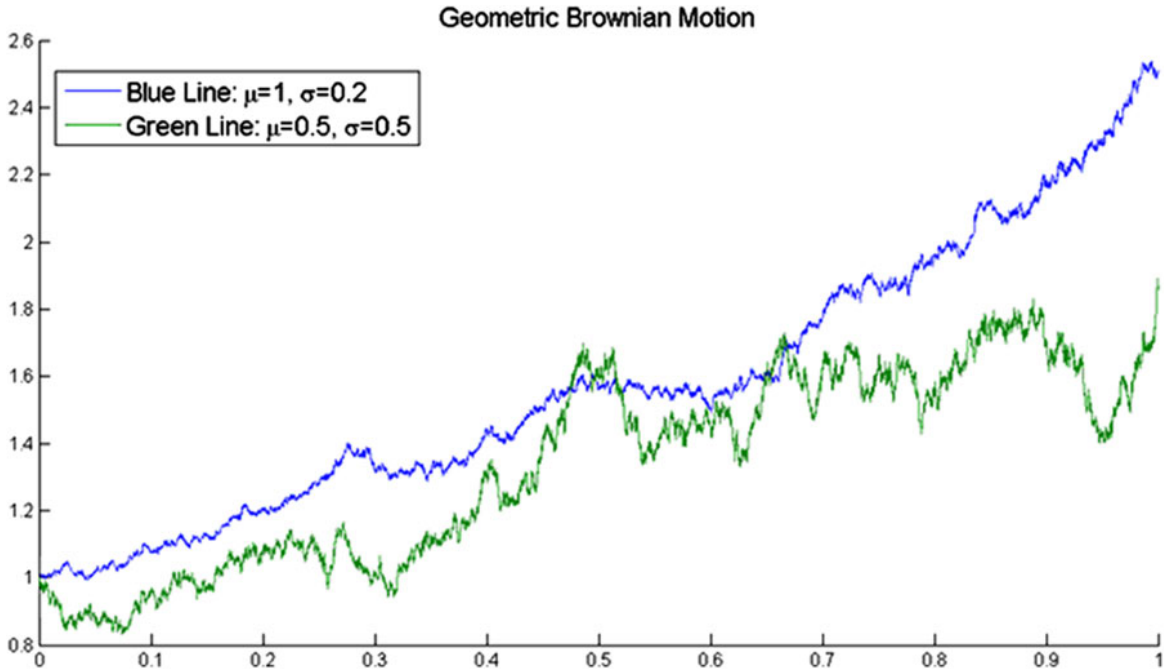

**Fig. 2.** GBM. Two particular realizations of the stochastic process called Geometric Brownian Motion (GBM) taken from the Wikipedia site http://en.wikipedia.org/wiki/Geometric_Brownian_motion. Their mean values are the exponential (7) with different values of *A* and *B* for each shown stochastic process.

In other words, mass extinctions that occurred in the past are indeed taken into account as unpredictable fluctuations in the number of living species that occurred in the particular realization of the GBM between −3.5 billion years and now. Hence, extinctions are 'unpredictable vertical downfalls' in that GBM plot that may indeed happen from time to time. Also notice that:
(1) The particular realization of GBM occurred over the last 3.5 billion years is very much unknown to us in its numerical details, but ...
(2) We would not care either, inasmuch as the theory of stochastic processes only cares about such statistical quantities like the mean value and the standard deviation curves, that are deterministic curves in time with known equations.

## GBM as the key to stochastic evolution of all kinds

### The N(t) GBM as stochastic evolution

On 8 January, 2012, this author came to realize that his Statistical Drake Equation, previously described is the special static case (i.e. 'the picture', so as to say) of a more general time-dependent statistical Drake equation (i.e. 'the movie', so as to say) that we study in this section. In other words, this result is a powerful generalization in time of all results described in sections: 'SETI and Darwinian Evolution merged mathematically' and 'GBM as the key to stochastic evolution of all kinds'. This section is thus an introduction to a new, exciting mathematical model that one may call 'Exponential Evolution in Time of the Statistical Drake Equation'.

To be precise, the number *N* in the statistical Drake equation (1), yielding the number of extraterrestrial civilizations now existing and communicating in the Galaxy, is replaced in this section by a stochastic process $N(t)$, jumping up and down in time like the number *e* raised to a Brownian motion, but actually in such a way that its mean value keeps increasing exponentially in time as

$$\langle N(t) \rangle = N_0\, e^{\mu t}. \qquad (15)$$

In (15), $N_0$ and µ are two constants with respect to the time variable *t*. Their meaning is, respectively:
(1) $N_0$ is the number of ET communicating civilizations at time $t=0$, namely 'now', if one decides to regard the positive times ($t>0$) as the future history of the Galaxy ahead of us, and the negative times ($t<0$) as the past history of the Galaxy.
(2) µ is a positive (if the number of ET civilizations increases in time) or negative (if the number of ET civilizations decreases in time) parameter that we may call 'the drift'. To fix the ideas, and to be optimistic, we shall suppose µ > 0.

This evolution in time of $N(t)$ is just what we expect to happen in the Galaxy, where the overall number $N(t)$ of ET civilizations does probably increase (rather than decrease) in time because of the obvious technological evolution of each civilization. However, this *N(t)* scenario is a stochastic one, rather than a deterministic one, and certainly does not exclude temporary setbacks, like the end of civilizations due to causes as diverse as:

(a) asteroid and comet impacts,
(b) rogue planets or stars, arriving from somewhere and disrupting the gravitational stability of the planetary system,
(c) supernova explosions that would 'fry' the entire nearby ET civilizations (think of AGN, the Active Nucleus Galaxies and ask: how many ET civilizations are dying in those galaxies right now?),
(b) ET nuclear wars, and
(e) possibly more causes of civilization end that we do not know about yet.

Table 2. *Summary of the properties of lognormal distribution that applies to the stochastic process N(t) = exponentially increasing number of ET communicating civilizations in the Galaxy, as well as the number of living species on earth over the last 3.5 billion years. Clearly, these two different GBM stochastic processes have different numerical values of $N_0$, μ and σ, but the equations are the same for both processes*

| | |
|---|---|
| Stochastic process | $N(t) = \begin{cases} 1) \text{ Number of ET Civilizations (in SETI).} \\ 2) \text{ Number of Living Species (in Evolution).} \end{cases}$ |
| Probability distribution | Lognormal distribution of the GBM |
| pdf | $N(t)\_pdf(n, N_0, \mu, \sigma, t) = \frac{1}{\sqrt{2\pi}\sigma\sqrt{t}\, n} e^{-\frac{\left[\ln(n) - \left(\ln N_0 + \mu t - \frac{\sigma^2 t}{2}\right)\right]^2}{2\sigma^2 t}}$ for $n \geqslant 0$ |
| Mean value | $\langle N(t) \rangle = N_0 \, e^{\mu t}$ |
| Variance | $\sigma^2_{N(t)} = N_0^2 \, e^{2\mu t} (e^{\sigma^2 t} - 1)$ |
| Standard deviation | $\sigma_{N(t)} = N_0 \, e^{\mu t} \sqrt{e^{\sigma^2 t} - 1}$ |
| All the moments, i.e. *k*th moment | $\langle N^k(t) \rangle = N_0^k \, e^{k\mu t} e^{(k^2 - k)\frac{\sigma^2 t}{2}}$ |
| Mode (= abscissa of the lognormal peak) | $n_{\text{mode}} \equiv n_{\text{peak}} = N_0 \, e^{\mu t} \, e^{-\frac{3\sigma^2 t}{2}}$ |
| Value of the mode peak | $f_{N(t)}(n_{\text{mode}}) = \frac{1}{N_0 \sqrt{2\pi} \sigma \sqrt{t}} e^{-\mu t} e^{\sigma^2 t}$ |
| Median (= fifty–fifty probability value for *N(t)*) | $\text{median} = m = N_0 \, e^{\mu t} e^{-\frac{\sigma^2 t}{2}}$ |
| Skewness | $\frac{K_3}{(K_2)^{\frac{3}{2}}} = (e^{\sigma^2 t} + 2)\sqrt{e^{\sigma^2 t} - 1}$ |
| Kurtosis | $\frac{K_4}{(K_2)^2} = e^{4\sigma^2 t} + 2e^{3\sigma^2 t} + 3e^{2\sigma^2 t} - 6$ |

Mathematically, we came to define the pdf of this exponentially increasing stochastic process *N(t)* as the lognormal

$$N(t)\_pdf(n, N_0, \mu, \sigma, t)$$
$$= \frac{1}{\sqrt{2\pi}\sigma\sqrt{t}\, n} e^{-\frac{\left[\ln(n) - \left(\ln N_0 + \mu t - \frac{\sigma^2 t}{2}\right)\right]^2}{2\sigma^2 t}} \quad \text{for} \quad 0 \leqslant n \leqslant \infty. \quad (16)$$

It is easy to prove that this lognormal pdf obviously fulfills the normalization condition

$$\int_0^\infty N(t)\_pdf(n, N_0, \mu, \sigma, t) \, dn$$
$$= \int_0^\infty \frac{1}{\sqrt{2\pi}\sigma\sqrt{t}\, n} e^{-\frac{\left[\ln(n) - \left(\ln N_0 + \mu t - \frac{\sigma^2 t}{2}\right)\right]^2}{2\sigma^2 t}} \, dn = 1. \quad (17)$$

Also, the mean value of (16) indeed yields the exponential curve (15)

$$\int_0^\infty n \cdot N(t)\_pdf(n, N_0, \mu, \sigma, t) \, dn$$
$$= \int_0^\infty n \frac{1}{\sqrt{2\pi}\sigma\sqrt{t}\, n} e^{-\frac{\left[\ln(n) - \left(\ln N_0 + \mu t - \frac{\sigma^2 t}{2}\right)\right]^2}{2\sigma^2 t}} \, dn = N_0 \, e^{\mu t}. \quad (18)$$

The proof of (17) and (18) is given in Appendix 11.A as the Maxima file 'GBM_as_N_of_t_v33' of Maccone (2012).

Table 2 summarizes the main properties of GBM, namely of this *N(t)* stochastic process.

*Our statistical Drake equation is the static special case of N(t)*

In this section, we prove the crucial fact that the lognormal pdf of our Statistical Drake Equation given in Table 1 is just 'the picture' case of the more general exponentially growing stochastic process *N(t)* ('the movie') having the lognormal pdf (16) as given in Table 2. To make things neat, let us denote by the subscript 'GBM' for both the μ and σ appearing in (16). The latter thus takes the form:

$$N(t)\_pdf(n, N_0, \mu_{\text{GBM}}, \sigma_{\text{GBM}}, t) = \frac{1}{\sqrt{2\pi}\sigma_{\text{GBM}}\sqrt{t}\, n}$$
$$\times e^{-\frac{\left[\ln(n) - \left(\ln N_0 + \mu_{\text{GBM}} t - \frac{\sigma_{\text{GBM}}^2 t}{2}\right)\right]^2}{2\sigma_{\text{GBM}}^2 t}} \quad \text{for} \quad 0 \leqslant n \leqslant \infty. \quad (19)$$

Similarly, let us denote by the subscript 'Drake' for both μ and σ appearing in the lognormal pdf given in the third line of

Table 1 (this is also equation (1.B.56) of Maccone (2012)), namely the pdf of our statistical Drake equation:

$$\text{lognormal\_pdf\_of\_Statistical\_Drake\_Eq}(n, \mu_{\text{Drake}}, \sigma_{\text{Drake}})$$
$$= \frac{1}{\sqrt{2\pi}\,\sigma_{\text{Drake}}\,n} e^{-\frac{(\ln(n) - \mu_{Drake})^2}{2\,\sigma^2}} \quad \text{for} \quad 0 \leqslant n \leqslant \infty. \tag{20}$$

Now, a glance at (19) and (20) reveals that they can be made to coincide if and only if the two simultaneous equations hold

$$\begin{cases} \sigma_{\text{GBM}}^2 \, t = \sigma_{\text{Drake}}^2, \\ \ln N_0 + \mu_{\text{GBM}} \, t - \frac{\sigma_{\text{GBM}}^2 \, t}{2} = \mu_{\text{Drake}}. \end{cases} \tag{21}$$

On the other hand, when we pass (so as to say) 'from the movie to the picture', the two σ must be the same thing, and so must be the two μ, that is, one must have:

$$\begin{cases} \sigma_{\text{GBM}} = \sigma_{\text{Drake}} = \sigma, \\ \mu_{\text{GBM}} = \mu_{\text{Drake}} = \mu. \end{cases} \tag{22}$$

Checking thus the upper equation (22) against the upper equation (21), we are only left with

$$t = 1. \tag{23}$$

Hence, $t = 1$ is the correct numeric value of the time leading 'from the movie to the picture'. Replacing this into the lower equation (21), and keeping in mind the upper equation (22), the lower equation (21) becomes

$$\ln N_0 + \mu_{\text{GBM}} - \frac{\sigma^2}{2} = \mu_{\text{Drake}}. \tag{24}$$

Since the two μ also must be the same because of the lower equation (22), then (24) further reduces to

$$\ln N_0 - \frac{\sigma^2}{2} = 0, \tag{25}$$

that is

$$N_0 = e^{\frac{\sigma^2}{2}} \tag{26}$$

and the problem of 'passing from the movie to the picture' is completely solved.

In conclusion, we have proven the following 'movie to picture' theorem:

The stochastic process $N(t)$ reduces to the random variable $N$ if, and only if, one inserts

$$\begin{cases} t = 1, \\ \sigma_{\text{GBM}} = \sigma_{\text{Drake}} = \sigma, \\ \mu_{\text{GBM}} = \mu_{\text{Drake}} = \mu, \\ N_0 = e^{\frac{\sigma^2}{2}} \end{cases} \tag{27}$$

into the lognormal probability density (16) of the stochastic process $N(t)$.

### GBM as the key to mathematics of finance

But what is this $N(t)$ stochastic process reducing to the lognormal random variable $N$ in the static case? Well, $N(t)$ is no less than the famous GBM, of paramount importance in the mathematics of finance. In fact, in the so-called Black–Scholes models, $N(t)$ is related to the log return of the stock price. Huge amounts of money all over the world are handled at Stock Exchanges according to the mathematics of the stochastic process $N(t)$, that is differently denoted $S_t$ there ('S' from Stock). But we would not touch these topics here, since this paper is about Evolution and SETI, rather than about stocks.

We just content ourselves to have proven that the GBM used in the mathematics of finance is the same thing as the exponentially increasing process $N(t)$ yielding the number of communicating ET civilizations in the Galaxy!

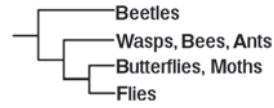

**Fig. 3.** A horizontal cladogram (taken from http://en.wikipedia.org/wiki/Cladogram) with the ancestor (not named) to the left.

### Darwinian Evolution re-defined as a GBM in the number of living species

*A concise introduction to cladistics and cladograms*

Cladistics is the science describing when new forms of life developed in the course of Evolution. Cladistics is thus the science of lineages, i.e. phylogenetic trees, like the one shown for instance in Figure 3, and it is today strongly based on computer codes, in turn based on high-level mathematics.

Our innovative contribution to cladistics and cladograms like the one in Figure 3 is to put the horizontal axis of time below them, and then realize that the cladograms branches are exponential functions of the time. In other words, these exponential arches are either increasing in time, or decreasing, or just staying constants (i.e. they are just horizontal lines, like the ones in Figure 3), but the length of these exponential arches is as long as the species they represent survived during the course of evolution.

This mathematical representation of the whole of evolution is:
(1) Easy, inasmuch as exponential functions like (7) are the easiest possible functions in mathematics.
(2) Clear, inasmuch as we know pretty well when a new species appeared in the course of evolution.
(3) GBM-based, inasmuch as the exponential arches indeed are the mean values in time of the corresponding 'unpredictable' GBMs yielding the number of members of that species living at a certain time in evolution. At last, the study of the mathematical properties of GBMs is now open to scientists, rather than only to bankers and businessmen, as it happened in the last 40 years (1973–2013). Note that, in 1997, the Nobel Prize in Economics was assigned to Robert C. Merton and Myron Scholes (Fischer Black had already died in 1995) for their mathematical discoveries (Black–Scholes–Merton models) based on GBMs. Perhaps, new Nobel Prizes will be assigned for applying GBMs to evolution and astrobiology.

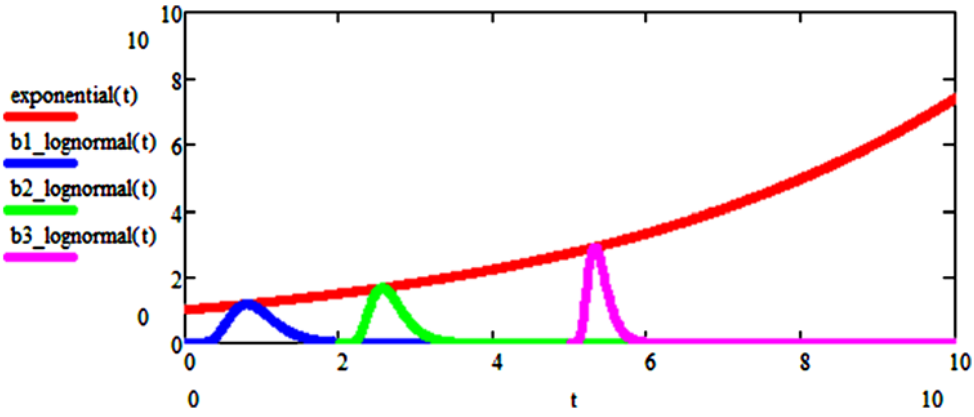

**Fig. 4.** Darwinian exponential as the envelope of *b*-lognormals. Each *b*-lognormal is a lognormal starting at a time (*t* = *b* = birth time) larger than zero and represents a different species 'born' at time *b* of Darwinian evolution.

*Cladistics: namely the GBM mean exponential as the locus of the peaks of b-lognormals representing each a different species started by evolution at time t = b > 0*

How is it possible to 'match' the GBM mean exponential curve with the lognormals appearing in the Statistical Drake Equation? Our answer to this question is 'by letting the GBM mean exponential become the envelope of the *b*-lognormals representing the cladistic branches, i.e. the new species that were produced by evolution at different times as long as evolution unfolded'. Let us now have a look at Figure 4.

The envelope shown in Figure 4 is not really an envelope in the strictly mathematical sense explained in calculus textbooks. However, it is 'nearly the same thing in practice' because it actually is the geometric locus of the peaks of all *b*-lognormals. We shall now explain this in detail.

First of all, let us write down the equation of the *b*-lognormal, i.e. of the lognormal starting at any positive instant *t* = *b* > 0 (while ordinary lognormals all start only at zero); in other words, (*t* − *b*) replaces *n* in the first equation in Table 1:

$$\begin{cases} \text{b\_lognormal}(t, \mu, \sigma, b) = \frac{1}{\sqrt{2\pi}\sigma \cdot (t-b)} e^{-\frac{(\ln(t-b)-\mu)^2}{2\sigma^2}} \\ \text{holding for } t > b \text{ and up to } t = \infty. \end{cases} \quad (28)$$

Then, notice that its peak falls at the abscissa *p* and ordinate *P* given by, respectively (as given by the 8th and 9th line in Table 1):

$$\begin{cases} p = b + e^{\mu-\sigma^2} = \text{b\_lognormal\_peak\_abscissa}, \\ P = \frac{e^{\frac{\sigma^2}{2}-\mu}}{\sqrt{2\pi}\sigma} = \text{b\_lognormal\_peak\_ordinate}. \end{cases} \quad (29)$$

Can we match the second equation (29) with the Darwinian exponential (7)? Yes, if we set at time *t* = *p*:

$$\begin{cases} A = \frac{1}{\sqrt{2\pi}\sigma}, \\ e^{Bp} = e^{\frac{\sigma^2}{2}-\mu}, \end{cases} \text{ that is } \begin{cases} A = \frac{1}{\sqrt{2\pi}\sigma}, \\ Bp = \frac{\sigma^2}{2} - \mu. \end{cases} \quad (30)$$

The last system of two equations may now be inverted, i.e. exactly solved with respect to μ and σ:

$$\begin{cases} \sigma = \frac{1}{\sqrt{2\pi}A}, \\ \mu = -Bp + \frac{1}{4\pi A^2}, \end{cases} \quad (31)$$

showing that each *b*-lognormal in Figure 4 (i.e. its μ and σ) is perfectly determined by the Darwinian exponential (namely by *A* and *B*) plus a precise value of the *b*-lognormal's peak time *p*. In other words, this is a one-parameter (the parameter is *p*) family of curves that are all constrained between the time axis and the Darwinian exponential.

Clearly, as long as one moves to higher values of *p*, the peaks of these curves become narrower and narrower and higher and higher, since the area under each *b*-lognormal always equals 1 (normalization condition).

*Cladogram branches are increasing, decreasing or stable (horizontal) exponential arches as functions of time*

It is now possible to understand how cladograms shape up in our mathematical theory of evolution: they depart from the time axis at birth time (*b*) of the new species and then either:

(1) Increase if the *b*-lognormal of the *i*th new species has (keeping in mind the convention $p_i < 0$ for past events, i.e. events prior to now):

$$\begin{cases} A_i = \frac{1}{\sqrt{2\pi}\sigma_i}, \\ B_i = \frac{\frac{\sigma_i^2}{2} - \mu_i}{p_i} > 0 \text{ that is } \mu_i > \frac{\sigma_i^2}{2}. \end{cases} \quad (32)$$

(2) Decrease if the same *b*-lognormal has (keeping in mind the convention $p_i < 0$ for past events):

$$\begin{cases} A_i = \frac{1}{\sqrt{2\pi}\sigma_i}, \\ B_i = \frac{\frac{\sigma_i^2}{2} - \mu_i}{p_i} < 0 \text{ that is } \mu_i < \frac{\sigma_i^2}{2}. \end{cases} \quad (33)$$

Table 3. *Summary of the statistical properties of the new random variable NoEv given by equation (35) and representing the stationary life of a new species born at time b and undergoing no evolution thereafter*

| Random variable | NoEv = NoEvolution probability = stationary life |
|---|---|
| Probability distribution | (no name yet) |
| pdf | $f_{\text{NoEv}}(t, \sigma, b) = \frac{1}{\sqrt{2\pi}\sigma\sqrt{t-b}} e^{-\frac{(\ln(t-b))^2}{2\sigma^2} - \frac{\sigma^2}{8}}$  $(t \geq b)$ |
| Mean value | $\langle \text{NoEv} \rangle = b + e^{\sigma^2}$ |
| Variance | $\sigma_{\text{NoEv}}^2 = e^{2\sigma^2}(e^{\sigma^2} - 1)$ |
| Standard deviation | $\sigma_{\text{NoEv}} = e^{\sigma^2}\sqrt{e^{\sigma^2} - 1}$ |
| Mode (= abscissa of the NoEv peak) | $t_{\text{mode}} \equiv t_{\text{peak}} = b + e^{-\frac{\sigma^2}{2}}$ |
| Value of the Mode Peak (= ordinate of the NoEv peak) | $f_{\text{NoEv}}(t_{\text{mode}}) = \frac{1}{\sqrt{2\pi}\,\sigma}$ |
| Median (= fifty–fifty probability value for NoEv) | $\text{median} = m = b + e^{\frac{\sigma^2}{2}}$ |
| Skewness | $\frac{K_3}{(K_2)^{\frac{3}{2}}} = (e^{\sigma^2} + 2)\sqrt{e^{\sigma^2} - 1}$ |
| Kurtosis | $\frac{K_4}{(K_2)^2} = e^{4\sigma^2} + 2e^{3\sigma^2} + 3e^{2\sigma^2} - 6$ |

(3) Keep staying constant (i.e. rather than exponential arches we have horizontal segments) for all time values for which the *i*th *b*-lognormal is characterized by:

$$\begin{cases} A_i = \frac{1}{\sqrt{2\pi}\sigma_i}, \\ B_i = 0 \text{ that is } \frac{\sigma_i^2}{2} = \mu_i. \end{cases} \quad (34)$$

This last case really is the most 'routine' one, inasmuch as the given species neither increases nor decreases in time, but rather, for generations and generations, 'the parents are born, mate, babies are born, the parents die, the babies mate, and so on endlessly'. This we call a stationary species. And, mathematically, the surprise is that a stationary species no longer is described by *b*-lognormals, but rather by the new probability density found by replacing the last equation (34) into (28), with the result that (28) becomes the new stationary pdf:

$$\text{stable\_pdf}(t, \sigma, b) = \frac{1}{\sqrt{2\pi}\sigma\sqrt{t-b}} e^{-\frac{(\ln(t-b))^2}{2\sigma^2}} e^{-\frac{\sigma^2}{8}}. \quad (35)$$

In plain words, this is the pdf for species undergoing no evolution at all, and this is not the pdf of the lognormal type because of the square root $\sqrt{t-b}$ appearing in (35) instead of $(t-b)$ appearing in (28). Clearly, more words and examples would be needed to better clarify our theory, but we have no space for that here. Table 3 yields the key statistical properties of stationary pdf (35) (see also Maccone (2011b)).

*KLT-Filtering in Hilbert space and Darwinian selection are "the same thing" in our theory…*

As a glance to the future developments of our mathematical theory of Darwinian evolution, let us now recall that the KLT is…a principal axes transformation in Hilbert space spanned by the eigenfunctions of the autocorrelation of a noise plus a possible signal in it. Put this way, the KLT (standing for Karhunen–Loève transform) may look 'hard to understand' (Maccone 2010c; Szumski 2011). However, we wish to describe by easy words that it amounts to the well-known Darwinian selection process. In fact, consider a Euclidean space with a large number *N* of dimensions. A point there means giving *N* coordinates. Each coordinate we assume to be 'a function of the body that Humans have in common with other animals, but other animals may or may not (because too primordial) have in common with humans. Then, the axis representing humans in this *N*-space has the largest variance of the set of points around it because humans have all functions. Monkeys have nearly the same number of functions as humans but in practice they have fewer of them. Thus, the Monkey axis in the *N*-space has the second largest variance around it. In the mathematical jargon of the KLT this is re-phrased by saying that humans are the dominant = first eigenvalue in the KLT of *N*-space, whereas Monkeys are the second eigenvalue, and so on for lower species, that are really almost 'noise' (i.e. rubbish) when compared with humans.

Now about filtering, i.e. extracting a tiny signal by virtue of the KLT from thick noise (this works so much better by virtue of the KLT than by virtue of the trivial FFT used by engineers all over the world, but that is another story, for which the reader may see Maccone (2010c)). Hence, just as Darwinian evolution filtered humans out of a lot of 'noise' (i.e. other lower level living organisms), so the KLT applied to the above large *N*-dimensional space may describe mathematically the selection carried on by Darwinian evolution across 3.5 billion years, but that requires another paper at least, or, better, the new book that this author is now writing.

*Conclusions about our statistical model for evolution and cladistics*

Evolution, as it occurred on Earth over the last 3.5 billion years, is only one chapter of the larger book encompassed by the Drake equation, which covers a time span of 10 billion years or so.

We sought to outline a unified and simple mathematical vision of both evolution and SETI, as the title of this paper says.

Our vision is based on the lognormal probability distribution characterizing $N$ in the statistical Drake equation.

We have shown that the envelope of such lognormal distributions 'changing in time' (b-lognormals) may account for the mean exponential increase of the number of living species on Earth over 3.5 billion years.

## Lifespans of living beings as b-lognormals

*Further extending b-lognormals as our model for all lifespans*

This section is devoted to explore the mathematical properties of b-lognormals. In fact, we shall now use them as the standard mathematical model to symbolize the lifespan of any living being, let this living being be a cell, or a human, or a society, or a human civilization, or even an ET civilization.

On the one hand, all such lifespans are of course finite in time, namely they are born at a certain instant $t = b$ ($b$ standing for 'birth') and they die at a later instant $t = d$ ($d$ standing for 'death'), with $d > b$.

On the other hand, b-lognormals like (28) are infinite in time, i.e. spanning from $t = b$ to $t = +\infty$, so one might immediately wonder how (28) might possibly represent a finite lifespan. Well, the answer to such a question will be given later in the next section, when we will introduce the notion of 'death instant' $t = d$ as the intersection point between the tangent to (28) in its descending inflexion point and the time axis. At the moment, we content ourselves with studying some mathematical properties of the infinite b-lognormal pdf (28).

This was done in a highly innovative editorial way in the author's book entitled 'Mathematical SETI', Maccone (2012). In fact, the mathematical proof of each of the theorems proven there was hardly demonstrated line-by-line in the text. On the contrary, the hardest calculations were performed by aid of Maxima, the powerful computer algebra code (also called Macsyma) created by NASA and MIT in the 1960s and now freely downloadable from the web site http://maxima.sourceforge.net/.

Hence, the reader may find them in the Maxima file 'b_lognormals_inflexion_points_and_DEATH_time.wmx' that is reprinted in Appendix 6.A. to the author's 2012 book. From now on, we shall simply state the equation numbers in that Maxima file proving a certain result about b-lognormals, and the interested reader will then find the relevant proof by reading the corresponding Maxima command lines ('i' = input lines) and output lines ('o' = output lines). This way of proving 'electronically' the mathematical results simplifies things greatly, if compared with the 'ordinary' lengthy proofs of traditional books, and students and researchers will be able to download for free the corresponding Maxima symbolic manipulator from the site: http://maxima.sourceforge.net/.

*Infinite b-lognormals*

Again, a b-lognormal simply is a lognormal pdf starting at any positive value $b > 0$ (called 'birth') rather than at the origin.

As such, a b-lognormal has the following equation in the independent variable $t$ (time) and with the three independent parameters $\mu$, $\sigma$ and $b$, of which $\mu$ is a real number, while both $\sigma$ and $b$ are positive numbers:

$$\begin{cases} \text{b\_lognormal}(t, \mu, \sigma, b) = \frac{1}{\sqrt{2\pi}\sigma(t-b)} e^{-\frac{(\ln(t-b)-\mu)^2}{2\sigma^2}} \\ \text{holding for } t > b \text{ and up to } t = \infty. \end{cases} \quad (36)$$

This we call the infinite b-lognormal, meaning that it extends to the right up to infinity. Its main mathematical properties are basically the same as those of the ordinary lognormals starting at zero and given in Table 1, with only one exception: all formulae representing an abscissa have the same expression as for ordinary lognormals with a $+b$ term added because of the right-shift of magnitude $b$. In other words, all infinite b-lognormals have the formulae given in the following Table 4 (a formal, analytical proof of all results in Table 4 can be found in Appendix 6.A of the author's book 'Mathematical SETI', Maccone (2012).

*From infinite to finite b-lognormals: defining the death time, d, as the time axis intercept of the b-lognormal tangent line at senility*

The b-lognormal extends up to $t = +\infty$ and this is in sharp contrast with the fact that every living being sooner or later dies at the finite time $d$ ('death') such that $0 < b < d < \infty$. We thus must somehow define this finite death time $d$ in order to let the b-lognormals become a realistic mathematical model for the life-and-death of every living being.

We solved this problem by defining the death time $t = d$ as the intercept point between the time axis and the straight line tangent to the b-lognormal at its descending inflexion point $t = s$, i.e. the tangent line to the lognormal curve at senility. And, from now on, we shall call finite b-lognormal any such truncated b-lognormal, ending just at $t = d$.

This section is devoted to the calculation of the equation yielding the $d$ point in terms of the b-lognormal's $\mu$ and $\sigma$, and the whole procedure is described at the lines %i45 thru %o56 of the file 'b_lognormals_inflexion_points_and_DEATH_time.wxm' in Appendix 6.B of Maccone (2012).

Let us start by recalling the simple formula yielding the equation of the straight line having an angular coefficient $m$ and tangent to the curve $y(t)$ at the point having the coordinates $(t_0, y_0)$:

$$y - y_0 = m(t - t_0). \quad (37)$$

Then, the value of $y_0$ clearly is the value of the b-lognormal at its senility time, given by the sixth line in Table 4, that is, rearranging:

$$y_0 = \frac{e^{\frac{\sigma\sqrt{\sigma^2+4}}{2}} e^{-\mu + \frac{\sigma^2}{4} - \frac{1}{2}}}{\sqrt{2\pi}\,\sigma}. \quad (38)$$

On the other hand, the abscissa of the senility time $t = s$ is given by the fifth line in Table 4, that is

$$t_0 = b + e^{\frac{\sigma\sqrt{\sigma^2+4}}{2} - \frac{3\sigma^2}{2} + \mu}. \quad (39)$$

Table 4. *Properties of the b-lognormal distribution, namely the infinite b-lognormal distribution given by (36). These are both statistical and geometric properties of the pdf (36), whose importance will become evident later*

| Probability distribution | b-lognormal, namely the infinite b-lognormal |
| --- | --- |
| pdf | $f_{b\text{-lognormal}}(t; \mu, \sigma, b) = \frac{1}{\sqrt{2\pi}\sigma} \cdot \frac{1}{(t-b)} e^{-\frac{(\ln(t-b)-\mu)^2}{2\sigma^2}}$ $(t \geqslant b \geqslant 0)$ |
| Abscissa of the ascending inflexion point | Adolescence $\equiv a = b + e^{-\frac{\sigma\sqrt{\sigma^2+4}}{2} - \frac{3\sigma^2}{2} + \mu}$ |
| Ordinate of the ascending inflexion point | $f_{b\text{-lognormal}}(\text{adolescence}) \equiv A = \frac{e^{\frac{\sigma\sqrt{\sigma^2+4}}{2}} e^{-\mu + \frac{\sigma^2}{4} - \frac{1}{2}}}{\sqrt{2\pi}\sigma}$ |
| Abscissa of the descending inflexion point | Senility $\equiv s = b + e^{\frac{\sigma\sqrt{\sigma^2+4}}{2} - \frac{3\sigma^2}{2} + \mu}$ |
| Ordinate of the descending inflexion point | $f_{b\text{-lognormal}}(\text{senility}) \equiv S = \frac{e^{-\frac{\sigma\sqrt{\sigma^2+4}}{2}} e^{-\mu + \frac{\sigma^2}{4} - \frac{1}{2}}}{\sqrt{2\pi}\sigma}$ |
| Mean value | $\langle b\_\text{lognormal}\rangle = b + e^\mu e^{\frac{\sigma^2}{2}}$ |
| Variance | $\sigma^2_{b\text{-lognormal}} = e^{2\mu} e^{\sigma^2}(e^{\sigma^2} - 1)$ |
| Standard deviation | $\sigma_{b\text{-lognormal}} = e^\mu e^{\frac{\sigma^2}{2}} \sqrt{e^{\sigma^2} - 1}$ |
| Peak Abscissa = mode | $b\text{-lognormal}_{peak} \equiv b\text{-lognormal}_{mode} \equiv p = b + e^\mu e^{-\sigma^2} = b + e^{\mu - \sigma^2}$ |
| Peak Ordinate = value of the mode peak | $f_{b\text{-lognormal}}(b\text{-lognormal}_{mode}) = \frac{1}{\sqrt{2\pi}\sigma} \cdot e^{-\mu} \cdot e^{\frac{\sigma^2}{2}} = \frac{1}{\sqrt{2\pi}\sigma} \cdot e^{\frac{\sigma^2}{2} - \mu}$ |
| Median (= fifty–fifty probability abscissa) | Median $= m = b + e^\mu$ |

Finally, we must find the expression of the angular coefficient *m* at the senility time, and this involves finding the *b*-lognormal's derivative at senility. Then Maxima has no problem to find this, and the lines %i48 and %o48 show that one gets, after some rearranging

$$m = -\frac{\sqrt{2} e^{\frac{7\sigma^2}{4}} (\sqrt{\sigma^2+4} - \sigma) e^{-\frac{\sigma\sqrt{\sigma^2+4} + 8\mu + 2}{4}}}{4\sqrt{\pi}\sigma^2}. \quad (40)$$

Inserting (38), (39) and (40) into (37) one obtains the equation of the desired straight line tangent to the *b*-lognormal at senility:

$$y - \frac{e^{\frac{\sigma\sqrt{\sigma^2+4}}{2}} e^{-\mu + \frac{\sigma^2}{4} - \frac{1}{2}}}{\sqrt{2\pi}\sigma} = -\frac{\sqrt{2} e^{\frac{7\sigma^2}{4}} (\sqrt{\sigma^2+4} - \sigma) e^{-\frac{\sigma\sqrt{\sigma^2+4} + 8\mu + 2}{4}}}{4\sqrt{\pi}\sigma^2}$$
$$\times \left(t - e^{\frac{\sigma\sqrt{\sigma^2+4}}{2} - \frac{3\sigma^2}{2} + \mu} - b\right). \quad (41)$$

In order to find the abscissa of the death point $t = d$, we just need to insert $y = 0$ into the above equation (41) and solve for the resulting *t*. Then Maxima yields at first a rather complicated result (%o52). However, keeping in mind that the term in *b* must obviously appear 'alone' in the final equation since the *b*-lognormal is only an ordinary lognormal shifted to make it start at *b*, the way to further simplify (41) becomes obvious, and the final result simply is

$$d = b + \frac{(\sqrt{\sigma^2+4} + \sigma)^2 e^{\frac{\sigma\sqrt{\sigma^2+4}}{2} - \frac{3\sigma^2}{2} + \mu}}{4}. \quad (42)$$

This is the 'death time' of all living beings born at any time $b > 0$.

*Terminology about various time instants related to a lifetime*

The reader is now asked to look carefully at Figure 5 to familiarize with mathematical notations and their meaning describing the lifetime of all living beings:

Obvious are the definitions of the instants of:
(1) birth (*b* = starting point on the time axis)
(2) adolescence (*a* = ascending inflexion abscissa, with ordinate A)
(3) peak (*p* = maximum point abscissa, with ordinate P)
(4) senility (*s* = descending inflexion abscissa, with ordinate S)
(5) and death (*d* = death abscissa = intercept between the time axis and the straight line tangent to the *b*-lognormal at the descending inflexion point).

*Terminology about various time spans related to a lifetime*

Also defined in Figure 5 are the obvious time segments called:
(1) Childhood ($C = a - b$)
(2) Youth ($Y = p - a$)
(3) Maturity ($M = s - p$),
(4) Decline ($D = d - s$),
(5) Fertility ($F = s - a$),
(6) Vitality ($V = s - b$)
(7) Lifetime ($L = d - b$).

Then, from all these definitions and from the mathematical properties of the *b*-lognormals listed in Table 4, one obtains immediately the following equations:

$$\text{Childhood} \equiv C = a - b = e^{-\frac{\sigma\sqrt{\sigma^2+4}}{2} - \frac{3\sigma^2}{2} + \mu}, \quad (43)$$

$$\text{Youth} \equiv Y = p - a = e^{\mu - \sigma^2} - e^{-\frac{\sigma\sqrt{\sigma^2+4}}{2} - \frac{3\sigma^2}{2} + \mu}, \quad (44)$$

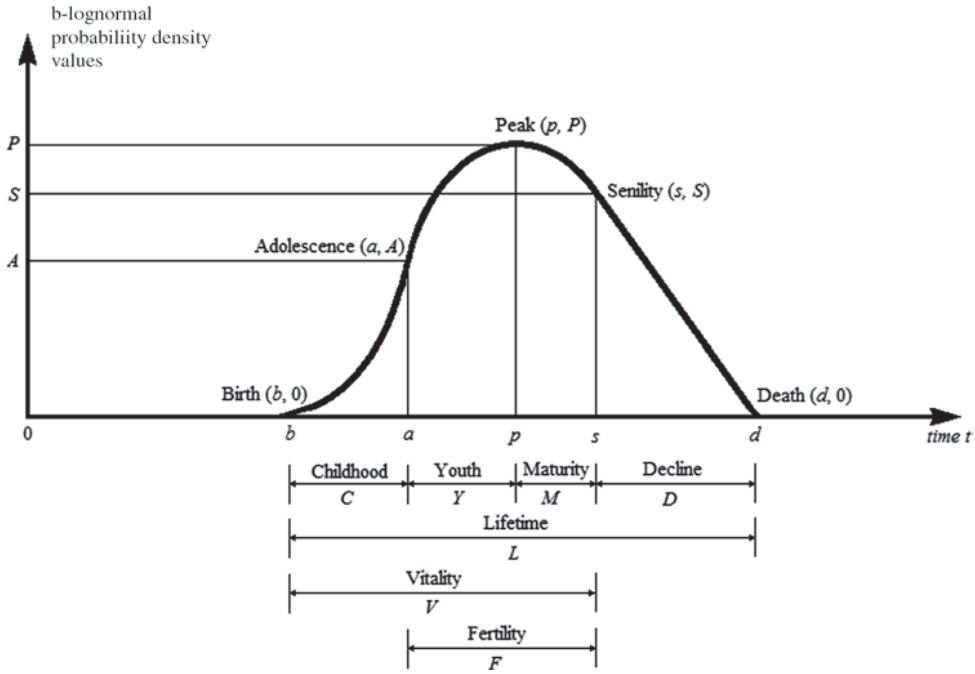

**Fig. 5.** Lifetime of all living beings, i.e. finite $b$-lognormal: definitions of the basic instants of birth ($b$ = starting point on the time axis), adolescence ($a$ = ascending inflexion abscissa, with ordinate $A$), peak ($p$ = maximum point abscissa, with ordinate $P$), senility ($s$ = descending inflexion abscissa, with ordinate $S$) and death ($d$ = death abscissa = intercept between the time axis and the straight line tangent to the $b$-lognormal at the descending inflexion point). Also defined are the obvious single-time-step-spanning segments called childhood ($C = a - b$), youth ($Y = p - a$), maturity ($M = s - p$), decline ($D = d - s$). In addition, also defined are the multiple-time-step-spanning segments of the all-covering lifetime ($L = d - b$), vitality ($V = s - b$) (i.e. lifetime minus decline) and fertility ($F = s - a$) (i.e. adolescence to senility).

$$\text{Maturity} \equiv M = s - p = e^{\frac{\sigma\sqrt{\sigma^2+4}}{2} - \frac{3\sigma^2}{2} + \mu} - e^{\mu - \sigma^2}. \quad (45)$$

$$\text{Decline} \equiv D = d - s = \frac{(\sqrt{\sigma^2+4} + \sigma)^2 e^{\frac{\sigma\sqrt{\sigma^2+4}}{2} - \frac{3\sigma^2}{2} + \mu}}{4} \quad (46)$$

$$- e^{\frac{\sigma\sqrt{\sigma^2+4}}{2} - \frac{3\sigma^2}{2} + \mu} = e^{\frac{\sigma\sqrt{\sigma^2+4}}{2} - \frac{3\sigma^2}{2} + \mu} \frac{\sigma(\sqrt{\sigma^2+4} + \sigma)}{2}.$$

$$\text{Fertility} \equiv F = s - a = e^{\frac{\sigma\sqrt{\sigma^2+4}}{2} - \frac{3\sigma^2}{2} + \mu} - e^{-\frac{\sigma\sqrt{\sigma^2+4}}{2} - \frac{3\sigma^2}{2} + \mu}$$

$$= 2 e^{-\frac{3\sigma^2}{2} + \mu} \sinh\left(\frac{\sigma\sqrt{\sigma^2+4}}{2}\right), \quad (47)$$

$$\text{Vitality} \equiv V = s - b = e^{\frac{\sigma\sqrt{\sigma^2+4}}{2} - \frac{3\sigma^2}{2} + \mu}, \quad (48)$$

$$\text{Lifetime} = L = d - b$$

$$= \frac{(\sqrt{\sigma^2+4} + \sigma)^2 e^{\frac{\sigma\sqrt{\sigma^2+4}}{2} - \frac{3\sigma^2}{2} + \mu}}{4}. \quad (49)$$

Obviously one also has

$$\text{Lifetime} = \text{Vitality} + \text{Decline} = (s - b) + (d - s)$$

$$= d - b = \frac{(\sqrt{\sigma^2+4} + \sigma)^2 e^{\frac{\sigma\sqrt{\sigma^2+4}}{2} - \frac{3\sigma^2}{2} + \mu}}{4}. \quad (50)$$

as one may check analytically by adding (48) and (46), and checking the result against (49).

In addition, dividing (46) by (48), all exponentials disappear and one obtains the important new equation

$$\frac{\text{Decline}}{\text{Vitality}} = \frac{\sigma(\sqrt{\sigma^2+4} + \sigma)}{2}. \quad (51)$$

This we shall use later in the section 'mathematical history of civilizations' in connection with the 'golden ratio' and 'golden $b$-lognormals'.

*Normalizing to one all the finite b-lognormals*

Finite $b$-lognormals are positive functions of time, as requested for any pdf, but they are not normalized to one yet, as it is also demanded for any pdf. This is because:

(1) If one computes the integral of the $b$-lognormal (36) between birth $b$ and senility $s$ one obtains

$$\int_b^s b\_\text{lognormal}(t, \mu, \sigma, b) dt$$

$$= \int_b^{e^{\frac{\sigma\sqrt{\sigma^2+4}}{2} - \frac{3\sigma^2}{2} + \mu}} \frac{1}{\sqrt{2\pi}\sigma(t-b)} e^{-\frac{(\ln(t-b)-\mu)^2}{2\sigma^2}} dt$$

$$= \frac{1}{2} + \frac{\text{erf}\left(\frac{\sqrt{2}}{4}\left(\sqrt{\sigma^2+4} - 3\sigma\right)\right)}{2}, \quad (52)$$

Table 5. *Finding the b-lognormal (i.e. finding both its µ and σ) given the birth time, b, and any two out of the four instants a=adolescence, p=peak, s=senility, d=death*

| Given | a | p | s | d |
|---|---|---|---|---|
| a | — | $\sigma = \dfrac{\ln\left(\frac{p-b}{a-b}\right)}{\sqrt{1+\ln\left(\frac{p-b}{a-b}\right)}}$  $\mu = \ln(a-b) + \dfrac{\sigma\sqrt{\sigma^2+4}}{2} + \dfrac{3\sigma^2}{2}$ | $\sigma = \sqrt{2}\sqrt{\sqrt{\left(\ln\left(\frac{\sqrt{s-b}}{\sqrt{a-b}}\right)\right)^2+1}-1}$  $\mu = \dfrac{\ln[(a-b)(s-b)]}{2} + \dfrac{3\sigma^2}{2}$ | No exact formula exists, only numeric approximations. |
| p | — | — | $\sigma = \dfrac{\ln\left(\frac{s-b}{p-b}\right)}{\sqrt{1-\ln\left(\frac{s-b}{p-b}\right)}}$  $\mu = \ln(s-b) - \dfrac{\sigma\sqrt{\sigma^2+4}}{2} + \dfrac{3\sigma^2}{2}$ | No exact formula exists, only numeric approximations. |
| s | — | — | — | $\sigma = \dfrac{d-s}{\sqrt{d-b}\sqrt{s-b}}$  $\mu = \ln(s-b) + \dfrac{2s^2+(-3d-b)s+d^2+bd}{(d-b)s-bd+b^2}$ History formulae |
| d | No exact formula exists, only numeric approximations. | No exact formula exists, only numeric approximations. | $\sigma = \dfrac{d-s}{\sqrt{d-b}\sqrt{s-b}}$  $\mu = \ln(s-b) + \dfrac{2s^2+(-3d-b)s+d^2+bd}{(d-b)s-bd+b^2}$ History formulae | — |

where erf(x) is the well-known error function of probability and statistics, defined by the integral

$$\text{erf}(x) = \frac{2}{\sqrt{\pi}} \int_0^x e^{-z^2} dz. \quad (53)$$

Notice that, during the integration in (52), the independent variable µ disappeared, leaving a result depending on σ only. We shall not prove (52) here: the proof can be found in Appendix 6.B of Maccone (2012), lines %i78 through%o79.

(2) If we add to (52) the integral of the descending straight line tangent to the b-lognormal at s, taken between s (given by the fifth line in Table 4) and d (given by (42)), we obtain

$$\int_s^d \text{y\_from\_eq.\_(36)} dt = \int_{b+e^{\frac{\sigma\sqrt{\sigma^2+4}}{2}-\frac{3\sigma^2}{2}+\mu}}^{b+\frac{\left(\sqrt{\sigma^2+4}+\sigma\right)^2 e^{\frac{\sigma\sqrt{\sigma^2+4}}{2}-\frac{3\sigma^2}{2}+\mu}}{4}}$$

$$\text{y\_from\_eq.\_(36)} dt = \frac{\left(\sqrt{\sigma^2+4}+\sigma\right) e^{\frac{3\sigma\sqrt{\sigma^2+4}}{4}-\frac{5\sigma^2}{4}-\frac{1}{2}}}{2^{\frac{5}{2}}\sqrt{\pi}}.$$

(54)

Once again µ disappeared, leaving a result depending on σ only. Again, we shall not prove (54) here: the proof can be found in Appendix 6.A of Maccone (2012), lines (%i85) through (%o87).

(3) In conclusion, adding (52) and (54), one gets the area under the finite b-lognormal (from b to d)

$$\text{Area\_under\_FINITE\_b-lognormal} = \int_b^d \text{FINITE\_b-lognormal} \, dt = K(\sigma) \quad (55)$$

with

$$K(\sigma) = \frac{1}{2} + \frac{\left(\sqrt{\sigma^2+4}+\sigma\right) e^{\frac{3\sigma\sqrt{\sigma^2+4}}{4}-\frac{5\sigma^2}{4}-\frac{1}{2}}}{2^{\frac{5}{2}}\sqrt{\pi}}$$

$$+ \frac{\text{erf}\left(\frac{\sqrt{2}}{4}\left(\sqrt{\sigma^2+4}-3\sigma\right)\right)}{2}. \quad (56)$$

In practice, it will be sufficient to compute the numeric value of K(σ) for a given σ and divide the corresponding finite b-lognormal by this value to have it normalized to one.

*Finding the b-lognormals given b and two out of the four a, p, s, d*

The question is now: having introduced the five points b, a, p, s, d, do some equations exist enabling one to determine the b-lognormal's µ and σ in terms of the birth time b (supposed to be always known) and any two more points out of the remaining four (a, p, s and d)? This author was able to discover several such pairs of equations, yielding µ and σ exactly (and not as numeric approximations) and they are all listed in Table 5. The mathematical proofs are given in Appendix 6.B of Maccone (2012), and will not be repeated here.

The most important out of all these equations are our brand-new history formulae, given by the two equations:

$$\begin{cases} \sigma = \dfrac{d-s}{\sqrt{d-b}\sqrt{s-b}}, \\ \mu = \ln(s-b) + \dfrac{2s^2 - (3d+b)s + d^2 + b\,d}{(d-b)s - b\,d + b^2}. \end{cases} \quad (57)$$

Essentially, these two equations allow us to find a $b$-lognormal when its birth, senility and death times are given. This is precisely what happens in the study of human history, since we certainly know when a past civilization was born (for instance when a new town was founded and later became the capital of a new empire), and when it died (because of war, usually). Less precisely we may know the time when its decline began (after reaching its peak), which is the $s$ appearing in the fifth line of Table 4. However, if one manages to find that out in history books, then the $b$-lognormal (36) is fully determined by our history formulae (57).

## Golden ratios and golden b-lognormals

### Is σ always smaller than 1?

So far, we have derived a number of properties of the $b$-lognormals given by (36) and representing the life of a living being. However, one question remains: is there any specific reason why σ should be smaller or larger than one? More precisely, while we know σ to be necessarily positive, no 'plausible' reason seems to exist for it to be smaller than one, as it appears to be numerically in majority of life forms.

To explore this topic a little more, consider a trivial rectangular triangle having catheti equal to 2 and σ, respectively. Owing to the well-known Pythagorean theorem, the hypotenuse obviously equals $\sqrt{\sigma^2 + 4}$. Since the hypotenuse always is longer than any of the catheti, we conclude that

$$\sqrt{\sigma^2 + 4} > \sigma. \quad (58)$$

Now insert (58) into (51). The result is

$$\frac{\text{Decline}}{\text{Vitality}} = \frac{\sigma(\sqrt{\sigma^2+4}+\sigma)}{2} > \frac{\sigma(\sigma+\sigma)}{2} = \frac{\sigma(2\sigma)}{2} = \frac{2\sigma^2}{2} = \sigma^2. \quad (59)$$

Since all variables in this inequality are positive, we may rewrite it as

$$\text{Decline} > \sigma^2 \cdot \text{Vitality}. \quad (60)$$

Now, in the majority of known life forms, it appears that the vitality time (i.e. the time between birth $b$ and senility $s$), i.e. $(s-b)$ is longer, or much longer than the decline time (i.e. the time between senility $s$ and death $d$, i.e. $(d-s)$). Thus, the only way to let (60) apply to biological reality is to conclude that it must be

$$\sigma^2 < 1 \quad \text{or} \quad \sigma^2 \ll 1 \quad (61)$$

from which one finally infers (for all life forms known to humans)

$$0 < \sigma < 1. \quad (62)$$

Actually, in Section 'Extrapolating history into the past: Aztecs' of this paper and in Chapter 7 of Maccone (2012) the numerical value of σ is estimated for $b$-lognormals of the most important historic Western Civilizations (Ancient Greece, Ancient Rome, Italian Renaissance, Portuguese Empire, Spanish Empire, French Empire, British Empire and finally American (USA) Empire), and in all cases the numerical value of σ turned out to be smaller than 1. This will be re-proven here in Section 'Extrapolating history into the past: Aztecs' as we just said.

Hence, one would tend to think that this $0 < \sigma < 1$ result must be a 'law of nature' of some kind, though we cannot offer any better proof.

There might, however, be 'pathological cases' of forms of life for which σ > 1 and so their decline would be larger or much larger than their vitality: just think of some science fiction movies like Star Wars, where some living being declares to be 900 years old or more…

Anyway, the dividing line between 'good' and 'bad' values of σ seems to be the σ = 1 case.

Is this case significant? Yes, very much, as we discover in the next section.

### Golden ratios and golden b-lognormals

If one lets σ = 1 into (51) one obtains

$$\frac{\text{Decline}}{\text{Vitality}} = \frac{\sigma(\sqrt{\sigma^2+4}+\sigma)}{2} = \frac{1(\sqrt{5}+1)}{2} = \frac{1+\sqrt{5}}{2}$$
$$= \text{golden ratio} = 1.6180339887\ldots = \phi. \quad (63)$$

This is the famous 'golden ratio', hailed by artists, architects and mathematicians as aesthetically pleasing for over 2000 years. In the Renaissance (1509), the Italian, Luca Pacioli (1445–1517) wrote a book about it by the Latin title of 'De Divina Proportione' (The Divine Proportion), with illustrations by Leonardo Da Vinci. Hence, let us go back to (63). We now wish to prove that the following 'divine proportion' holds among Lifetime, Vitality and Decline (but only for those life forms having σ = 1, of course):

$$\frac{\text{Lifetime}}{\text{Decline}} = \frac{\text{Decline}}{\text{Vitality}} = \text{Golden\_Ratio} \equiv \phi \equiv \frac{1+\sqrt{5}}{2}$$
$$= 1.618\ldots. \quad (64)$$

For the proof, admit for a moment that (64) holds good. Then, because of (50), the supposed (64) may be rewritten as

$$\phi = \frac{\text{Lifetime}}{\text{Decline}} = \frac{\text{Vitality} + \text{Decline}}{\text{Decline}} = \frac{\text{Vitality}}{\text{Decline}} + 1$$
$$= \frac{1}{\frac{\text{Decline}}{\text{Vitality}}} + 1 = \frac{1}{\phi} + 1. \quad (65)$$

Table 6. *Golden b-lognormal distribution, i.e. the b-lognormal having σ =1, and its statistical properties*

| Probability distribution | Golden *b*-lognormal |
| --- | --- |
| pdf | $f_{\text{Golden\_b-lognormal}}(t; \mu, b) = \frac{1}{\sqrt{2\pi}} \frac{1}{(t-b)} e^{-\frac{(\ln(t-b)-\mu)^2}{2}}$ $(t \geqslant b \geqslant 0)$ |
| Abscissa of the ascending inflexion point | Adolescence $\equiv a = b + e^{\mu - \frac{3+\sqrt{5}}{2}}$ |
| Ordinate of the ascending inflexion point | $f_{\text{Golden\_b-lognormal}}(\text{adolescence}) \equiv A = \frac{e^{-\mu - \frac{1+\sqrt{5}}{4}}}{\sqrt{2\pi}}$ |
| Abscissa of the descending inflexion point | Senility $\equiv s = b + e^{\mu + \frac{\sqrt{5}-3}{2}}$ |
| Ordinate of the descending inflexion point | $f_{\text{Golden\_b-lognormal}}(\text{Senility}) \equiv S = \frac{e^{-\mu - \frac{1-\sqrt{5}}{4}}}{\sqrt{2\pi}}$ |
| Abscissa of the death point | $d = b + \frac{(\sqrt{5}+1)^2 e^{\mu + \frac{\sqrt{5}-3}{2}}}{4}$ |
| Mean value | $\langle \text{Golden\_b-lognormal} \rangle = b + e^{\mu + \frac{1}{2}}$ |
| Variance | $\sigma^2_{\text{Golden\_b-lognormal}} = e^{2\mu + 1}(e - 1)$ |
| Standard deviation | $\sigma_{\text{Golden\_b-lognormal}} = e^{\mu + \frac{1}{2}}\sqrt{e - 1}$ |
| Peak Abscissa = mode | $\text{Golden\_b-lognormal}_{\text{peak}} \equiv \text{Golden\_b-lognormal}_{\text{mode}} \equiv p = b + e^{\mu - 1}$ |
| Peak ordinate = value of the mode peak | $f_{\text{Golden\_b-lognormal}}(\text{Golden\_b-lognormal}_{\text{mode}}) = \frac{1}{\sqrt{2\pi}} e^{\frac{1}{2} - \mu}$ |
| Median (= fifty–fifty probability value) | Median $= m = b + e^{\mu}$ |
| Skewness | $\frac{K_3}{(K_2)^{\frac{3}{2}}} = (e + 2)\sqrt{e - 1} = 6.185...$ |
| Kurtosis | $\frac{K_4}{(K_2)^2} = e^4 + 2e^3 + 3e^2 - 6 = 110.936...$ |

Thus, if this is correct, we reach the conclusion that φ must fulfill the equation

$$\phi = \frac{1}{\phi} + 1 \quad \text{that is} \quad \phi^2 - \phi - 1 = 0. \tag{66}$$

Solving this quadratic equation in φ yields

$$\phi = \frac{-(-1) \pm \sqrt{1 - 4(-1)}}{2} = \frac{1 \pm \sqrt{5}}{2}. \tag{67}$$

Discarding the negative root in (67) (since the ratio of positive quantities may only yield a new positive quantity) leaves the positive root only, and this is just the golden ratio appearing in (63). Thus, we met with no contradiction in assuming the 'divine proportion' (64) to be true, and so it is true indeed.

As a consequence, it appears quite natural to call *golden b-lognormal* the particular case σ = 1 of (36), that is

$$\begin{cases} \text{golden\_b\_lognormal}(t, \mu, b) = \frac{1}{\sqrt{2\pi}(t-b)} e^{-\frac{(\ln(t-b)-\mu)^2}{2}} \\ \text{holding for } t > b \text{ and up to } t = \infty. \end{cases} \tag{68}$$

This is a 'new' statistical distribution, whose main statistical properties are listed in Table 5, of course derived by setting σ = 1 into the corresponding entries of Table 4.

Actually, rather than being only a single curve, (68) is a one-parameter family of curves in the (*t*, Golden_b-lognormal) plane, the parameter being μ. One is thus led to wonder what properties might this family of Golden *b*-lognormals possibly have. Then, this author discovered a simple theorem: all the golden *b*-lognormals (68) have their peaks lying on the equilateral hyperbola of equation

$$\text{Golden\_b-lognormal\_PEAK\_LOCUS}(t, b) = \frac{1}{\sqrt{2\pi}\sqrt{e}(t-b)}. \tag{69}$$

The proof is easy: just solve for μ the equation (line 11 in Table 5) yielding the peak abscissa of (68). The result is

$$\mu = \ln(p - b) + 1. \tag{70}$$

Inserting (70) into the expression for the peak height *P* given in line 12 of Table 5, (69) is found, and the theorem is thus proven. Figure 6 shows immediately the equilateral hyperbola (69) for the case *b* = 2.

But all these considerations about the golden ratio and the golden *b*-lognormals appear to be only an iceberg's tip if one thinks of the many known results relating the golden ratio to the Fibonacci numbers, Lucas numbers, and so on. Hence,

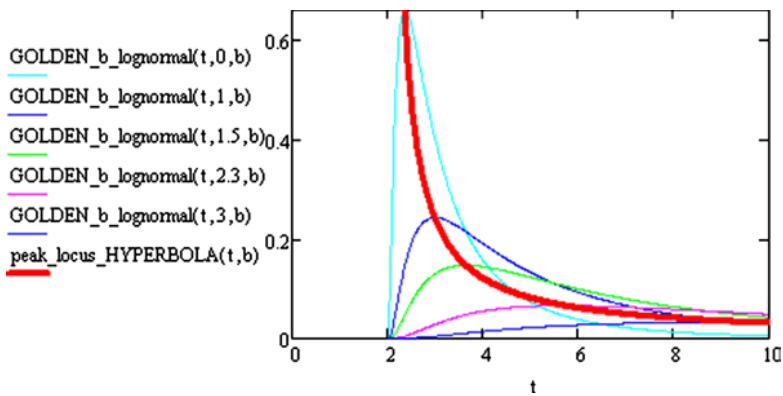

**Fig. 6.** The geometric locus of the peaks of all golden $b$-lognormals (in the above diagram starting all at $b=2$) as the parameter μ takes on all positive values ($0 \leq \mu \leq \infty$), is the equilateral hyperbola given by (69).

Table 7. *Finding the b-lognormals of eight among the most important civilizations of the Western world: Ancient Greece, Ancient Rome, Renaissance Italy, Portugal, Spain, France, Britain and the USA. For each such civilization three input dates are assigned on the basis of historic facts: (1) the birth time, b; (2) the senility time, s, i.e. the time when the decline began, and (3) the death time, d, when the civilization reached a formal end. From these three inputs and the two equations (57) the b-lognormal of each civilization may be computed. As a result, that civilization's peak is found, as shown in the last two columns. In general, this peak time turns out to be in agreement with the main historical facts*

|  | $b$ = Birth time | $s$ = Senility time | $d$ = Death time | $p$ = Peak time | $p$ = Peak ordinate |
|---|---|---|---|---|---|
| Ancient Greece | 600 BC Mediterranean Greek coastal expansion. | 323 BC Alexander the Great's death. Hellenism starts. | 30 BC Cleopatra's death: last Hellenistic queen. | 434 BC Pericles' Age. Democracy peak. Arts and science peak. | $2.488 \times 10^{-3}$ |
| Ancient Rome | 753 BC Rome founded. Italy seized by Romans by 270 BC. | 235 AD Military Anarchy starts. Rome not capital any more. | 476 AD Western Roman Empire ends. Dark Ages start. | 59 AD Christianity preached in Rome by Saints Peter and Paul against slavery. | $2.193 \times 10^{-3}$ |
| Renaissance Italy | 1250 Frederick II dies. Middle Ages end. Free Italian towns. | 1564 Council of Trent. Tough Catholic and Spanish rule. | 1660 1600 Bruno burned, 1642 Galileo dies. 1667 Cimento Academy Shut. | 1497 Renaissance art and architecture. Science. Copernican revolution. | $5.749 \times 10^{-3}$ |
| Portugal | 1419 Madeira island discovered. | 1822 Brazil independent, colonies retained. | 1999 Last colony Macau lost. | 1716 Black slave trade to Brazil at its peak. | $3.431 \times 10^{-3}$ |
| Spain | 1492 Columbus discovers America. | 1805 Spanish fleet lost at Trafalgar. | 1898 Last colonies lost to the USA. | 1741 California to be settled by Spain, 1759–76. | $5.938 \times 10^{-3}$ |
| France | 1524 Verrazano first in New York bay. | 1815 Napoleon defeated at Waterloo. | 1962 Algeria lost, as most colonies. | 1732 French Canada and India conquest tried. | $4.279 \times 10^{-3}$ |
| Britain | 1588 Spanish Armada Defeated. | 1914 World War One won at a high cost. | 1973 The UK joins European EEC. | 1868 Victorian Age. Science: Faraday and Maxwell. | $8.447 \times 10^{-3}$ |
| USA | 1898 Philippines, Cuba, Puerto Rico seized. | 2001 9/11 terrorist attacks. | 2050? Will the USA yield to China? | 1973 Moon landings, 1969–72. | 0.013 |

much more work is needed certainly in this new field we have uncovered.

## Mathematical history of civilizations

### Civilizations unfolding in time as b-lognormals

Centuries of human history on Earth should have taught us something.

Basically, civilizations are born, fight against each other and 'die', merging, however, with newer civilizations.

To cast all this in terms of mathematical equations is hard. The reason nobody has done so is because the task is so daunting. Indeed, no course on 'Mathematical History' is taught at any university in the world.

In this section, we will have a stab at this. Our idea is simple: any civilization is born, reaches a peak, then declines…just like a $b$-lognormal!

### Eight examples of western historic civilizations as finite b-lognormals

We now offer eight examples of such a view: the historic development of the civilizations of:

(1) Ancient Greece
(2) Ancient Rome
(3) Renaissance Italy
(4) Portuguese Empire
(5) Spanish Empire
(6) French Empire

(7) British Empire
(8) American (USA) Empire

Other historic empires (for instance the Dutch, German, Russian, Chinese, and Japanese ones, not to mention the Aztec and Incas Empires, or the Ancient ones, like the Egyptian, Persian, Parthian, or the medieval Mongol Empire) should certainly be added to such a picture, but we regret we do not have the time to carry on those studies in this paper. Those historic-mathematical studies will be made at a later stage of development of this new research field that, in our view, is 'Mathematical History': the mathematical view of human history based on *b*-lognormal probability distributions.

To summarize this section's content, for each one of the eight civilizations listed above, we define:

(a) Birth *b*, namely the year when that civilization was supposed to be 'born', even if only approximately in time.
(b) Senility *s*, namely the year of an historic event that marked the beginning of the decline of that civilization.
(c) Death *d*, namely the year when an historic event marked the 'official passing away' of that civilization from history.

Then, consider the two equations (57). For each civilization, these two equations allow us to compute both μ and σ in terms of the three assigned numbers (*b, s, d*). As a consequence, the time of the given civilization peak is found immediately from the upper equation (29), that is

peak_time = abscissa_of_the_maximum = *p*

$$= b + e^{\mu - \sigma^2}. \quad (71)$$

Also, we can then write down the equation of the corresponding *b*-lognormal immediately. The plot of this function of time gives a clear picture of the historic development of that civilization, though, to save space, we prefer not to reproduce here the above eight *b*-lognormals separately.

Inserting the peak time (71) into (36), the peak ordinate of the civilization is found, namely 'how civilized that civilization was at its peak,' and this is explicitly given by the lower equation (29), namely:

$$\text{peak\_ordinate} = P = \frac{e^{\frac{\sigma^2}{2} - \mu}}{\sqrt{2\pi}\sigma}. \quad (72)$$

Table 7 summarizes the three input data (*b, s, d*) drawn by the author from history textbooks, and then the two output data (*p, P*) of that Civilization's peak, namely its best legacy to other subsequent Civilizations.

*Plotting all b-lognormals together and finding the trends*

Having determined the *b*-lognormal for each civilization we wish to study, the time is ripe to plot all of them together and 'see what the trends are'. This is done in Figure 7.

We immediately notice some trends:

(1) The first two civilizations in time (Greece and Rome) are separated from the six modern ones by a large, 1000 years gap. This is of course the Middle Ages, i.e. the Dark Ages that hampered the development of Western Civilization for about 1000 years. Carl Sagan said, 'the millennium gap in the middle of the diagram represents a poignant lost opportunity for the human species', Sagan (1980).

(2) While the first two civilizations of Greece and Rome lasted more than 600 years each, all modern civilizations lasted much less: 500 years at most, but really less, or much less indeed.

(3) Since *b*-lognormals are pdfs, the area under each *b*-lognormal must be the same, i.e. just 1 (normalization condition). Thus, the shorter a civilization lives, the highest its peak must be! This is obvious from Figure 7: Greece and Rome lasted so long, and their peak was so much smaller than the British or the American peak!

(4) In other words, our theory accounts for the 'higher level of the more recent historic civilizations' in a natural fashion, with no need to introduce further free parameters. Not a small result, we think.

(5) All these remarks lead to the Appendix 7.A file in Maccone (2012) and the Figures therewith, starting with Figure 7 hereafter.

*b-lognormals of alien civilizations*

So much about the past. But what about the future ? What are the *b*-lognormals of ET civilizations in this Galaxy?

Nobody knows, of course. And nobody will know as long as the SETI scientists are unable to detect the first signs of an extraterrestrial civilization. Science fiction fans, however, might take pleasure in casting the Star Trek timeline into the mathematical language of *b*-lognormals. In this regard, interesting is 'The Star Trek Chronology', by Okuda & Okuda (1996). Also, the interested readers should get a copy of the great book by Finney & Jones (1986). This book is 'revolutionary', inasmuch as it re-reads the history of many human past civilizations with the glasses of the new science of SETI. We learned a lot from this book, but ... no mathematics is there, just words. Our achievement was to "convert that book into equations".

**Extrapolating history into the past: Aztecs**

*Aztecs–Spaniards as an example of two suddenly clashing civilizations with large technology gap*

The only example we know for sure about two suddenly clashing civilizations with very different technological levels comes from human history. That was in 1519, when the Spaniard Hernán Cortés, with 600 men, 15 horsemen, 15 cannons and hundreds of indigenous carriers and warriors, was able to subdue the Aztec empire of Montezuma II, numbering some 20 million people. How was that possible?

Well, we claim that basically there was a psychological breakdown in the Aztecs due to their obvious technological inferiority to the Spaniards, causing the Aztecs to regard the Spaniards as 'Semi-Gods', or 'Gods'. We also claim that this is precisely what might happen to humans when they meet for the first time with a much more technologically advanced alien civilization in the Galaxy: humans might be shocked and paralysed by Alien superiority, thus simply surrendering to alien will.

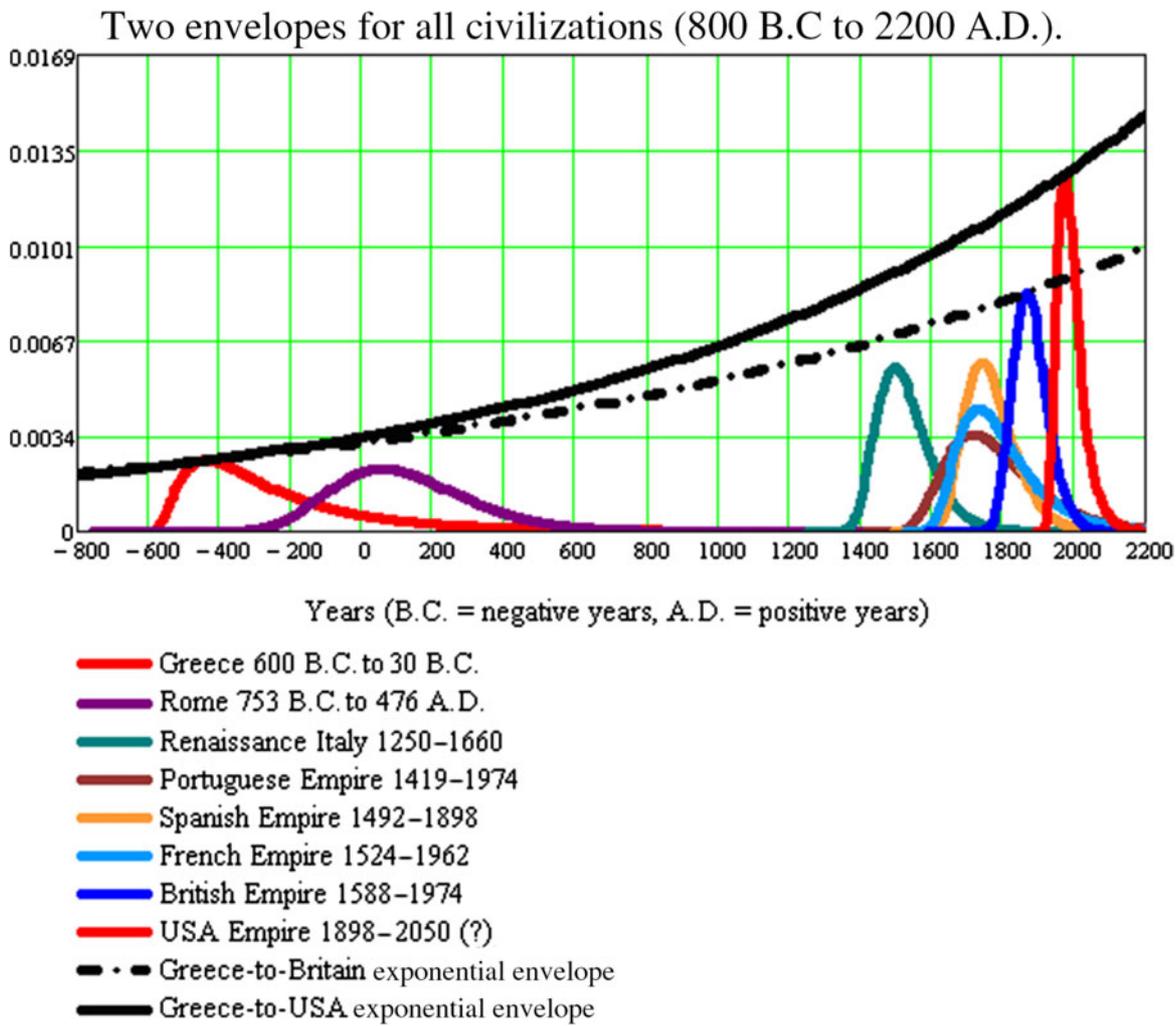

**Fig. 7.** Showing the *b*-lognormals of eight civilizations in Western history, with two exponential envelopes for them.

We also claim, however, that this human–alien sudden clash might be somehow softened were humans able to make a mathematical estimate of how much more advanced than us aliens will be. This mathematical theory of the technological civilization level is now developed in this section with a reference to the Aztecs–Spaniards example.

*'Virtual Aztecs' method to find the 'True Aztecs' b-lognormal*

First of all, this author has developed a mathematical procedure to correctly locate the *b*-lognormal of past human civilizations in time.

Consider the Aztec–Spaniard case: how much were the Spaniards more technologically developed than the Aztecs? Well, we claim that the answer to this question comes from the consideration of wheels. The use of wheels was unknown to the Aztecs. However, although they did not develop the wheel proper, the Olmec and certain other western hemisphere cultures seem to have approached it, as wheel-like worked stones have been found on objects identified as children's toys. This is just the point: we assume that the Aztecs 'were on the verge' of discovering wheels when the Spaniards arrived in 1519.

But then, when had wheels been discovered by the Asian–European civilizations?

Evidence of wheeled vehicles appears from the mid-4th millennium BC, near-simultaneously in Mesopotamia, the Northern Caucasus (Maykop culture) and Central Europe, so that the question of which culture originally invented the wheeled vehicle remains unresolved and under debate.

The earliest well-dated depiction of a wheeled vehicle (a wagon–four wheels, two axles), is on the Bronocic pot, a *ca.* 3500–3350 B.C. clay pot excavated in a Funnelbeaker culture settlement in southern Poland. The wheeled vehicle spread from the area of its first occurrence (Mesopotamia, Caucasus, Balkans, Central Europe) across Eurasia, reaching the Indus Valley by the 3rd millennium BC. During the 2nd millennium BC, the spoke-wheeled chariot spread at an increased pace, reaching both China and Scandinavia by 1200 B.C. In China, the wheel was certainly present with the adoption of the chariot in *ca.* 1200 B.C.

To fix the numbers, we shall thus assume that wheels had been discovered by the Asian–Europeans about 3500 B.C. Hence, summing 3500 plus 1519 (when the wheel-less Aztecs clashed against the wheel-aware Spaniards), we obtain about 5000 years of technological difference of level among these two civilizations. And 5000 years means 50 centuries, and not just 'a few centuries' of Aztecs inferiority, as historians having no mathematical background have superficially claimed in the past: our *b*-lognormal theory is quantitatively much more precise than just 'words'! However, let us now extend into the past, up to 3800 B.C., the older diagram shown in Figure 7. Adding the Greece-to-Spain *b*-lognormal (shown in Figure 8), the newer, resulting diagram extending to 3800 B.C. is shown in Figure 9.

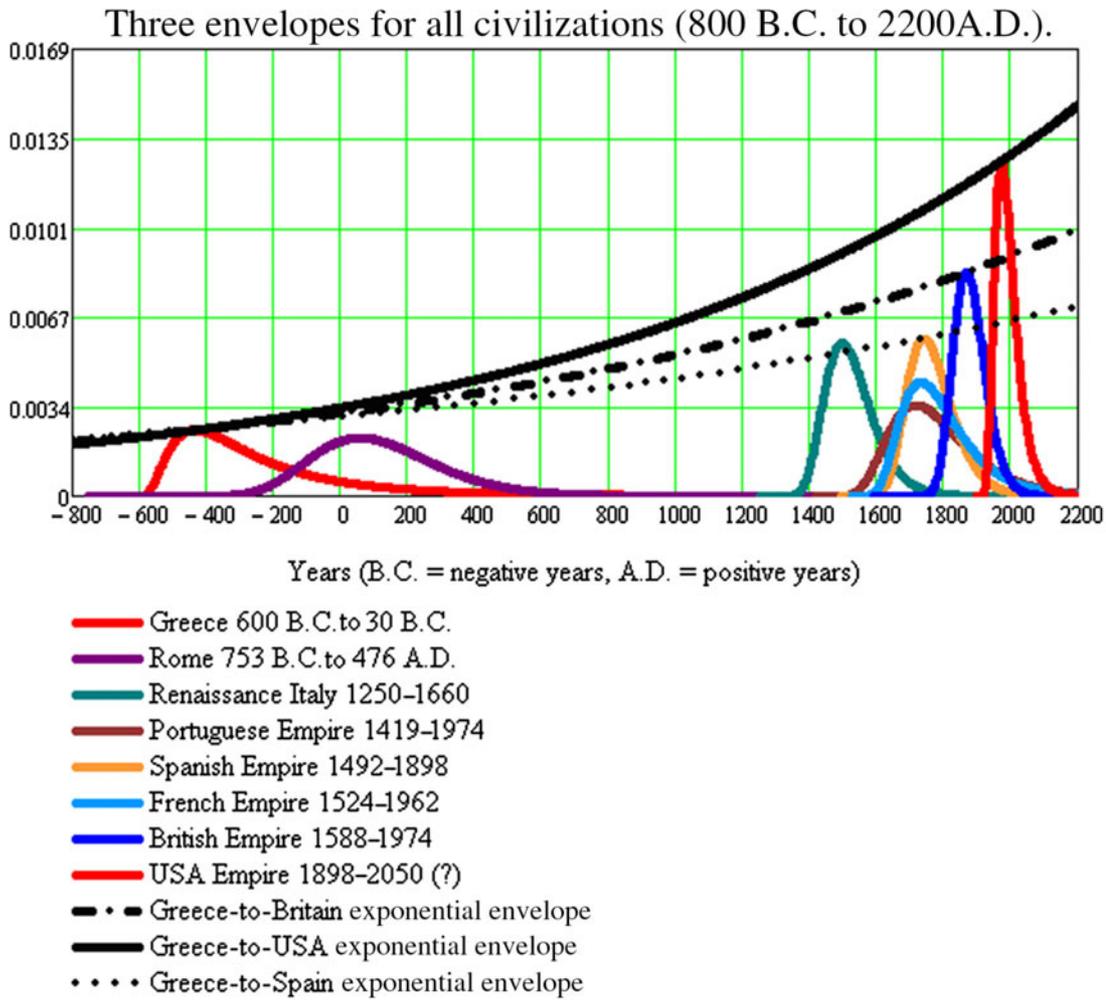

**Fig. 8.** Showing the *b*-lognormals of eight Western civilizations over the 5000 years (= 50 centuries) time span from 3800 B.C. to 2200 A.D. In addition, three exponential 'envelopes' (or, more precisely, three 'loci of the maxima') are shown:
(1) The Ancient-Greece-peak (434 BC) to Britain's peak (1868) exponential, namely the dash-dot black curve.
(2) The Ancient-Greece-peak (434 BC) to USA peak (1973) exponential, namely the solid black curve.
(3) The Ancient-Greece-peak (434 BC) to Spain peak (1741) exponential, namely the dot–dot black curve.
The Greece-to-Spain exponential was introduced since it is needed to understand the clash between the Aztecs and the Spaniards (1519–1521), as described by the 'Virtual Aztec' *b*-lognormal, going back 50 centuries before 1519 (see Figure 9).

In Figure 9, the virtual Aztec *b*-lognormal is the *b*-lognormal peaking at the time in the past when the Western civilizations discovered the wheel, i.e. about 3500 B.C. in Mesopotamia, Southern Caucasus and Central Europe. This *b*-lognormal is the dash–dash black curve in Figure 9. The Aztecs started their expansion in central Mexico in 1325, so when Cortez arrived in 1519 they were a civilization 1519–1325 = 194 years old.

Reporting this 194 years lapse before the year 3500 B.C., we find that the virtual Aztecs had been 'born' 194 years earlier, namely in 3694 BC, which is thus the *b*-value of the virtual Aztec *b*-lognormal

$$b_{VA} = -3694. \tag{73}$$

Then we have to find the *b*-lognornal itself, i.e. its $\mu_{VA}$ and $\sigma_{VA}$. In other words, we have to find $\mu_{VA}$ and $\sigma_{VA}$ knowing only the two peak coordinates, $p_{VA} = -3500$ and $P_{VA}$ (the numeric value of the peak height $P_{VA}$ is obviously known, since it equals the value of the Greece-to-Spain exponential, the dot–dot curve in Figure 9):

$$\begin{cases} -3500 = p_{VA} = b_{VA} + e^{\mu_{VA} - \sigma_{VA}^2}, \\ \text{Greece\_to\_Spain\_EXPONENTIAL} = \\ 7.305 \times 10^{-4} = P_{VA} = \dfrac{e^{\frac{\sigma_{VA}^2}{2} - \mu_{VA}}}{\sqrt{2\pi}\sigma_{VA}}. \end{cases} \tag{74}$$

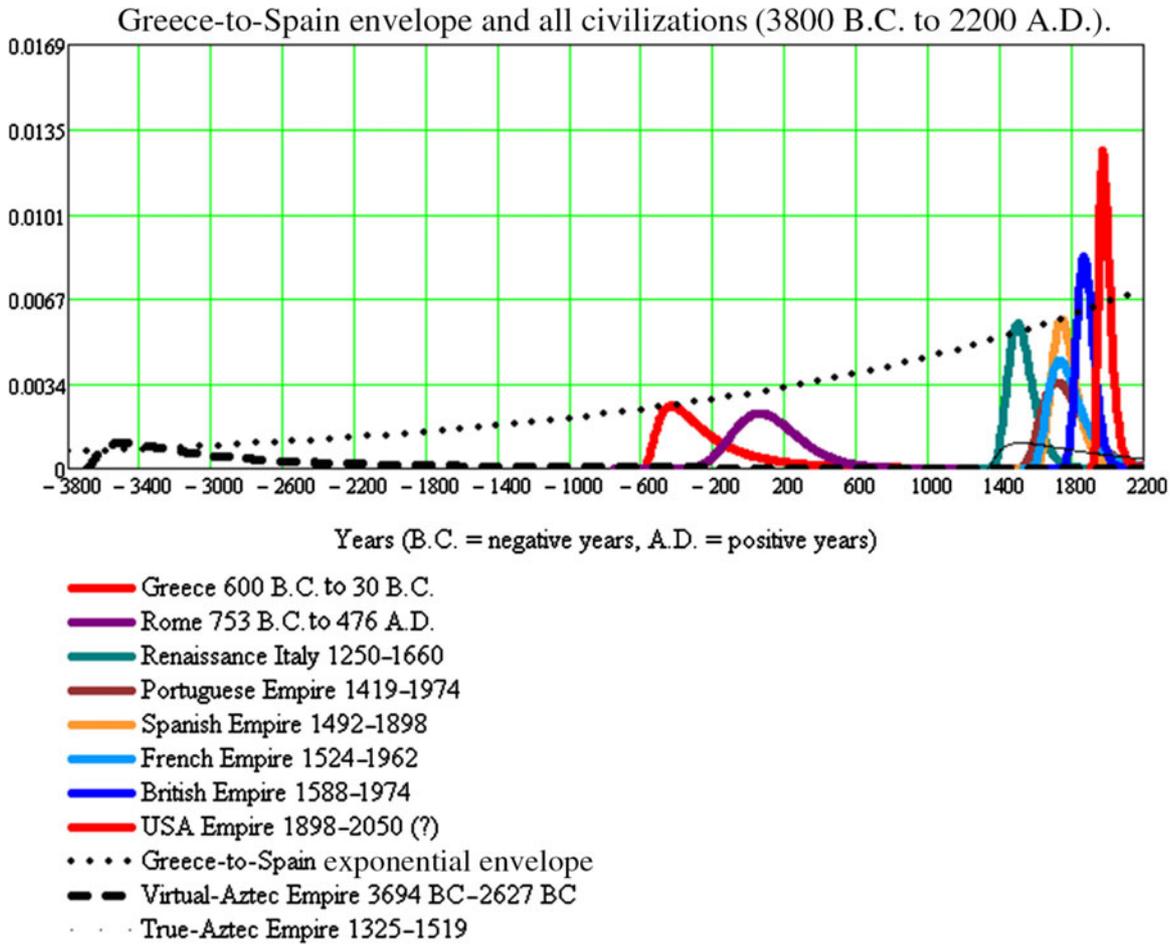

**Fig. 9.** The virtual Aztec $b$-lognormal is the $b$-lognormal peaking at a time in the past when the Western civilizations discovered the wheel, i.e. about 3500 B.C. in Mesopotamia, Southern Caucasus and Central Europe. This $b$-lognormal is the dash–dash black curve in the above diagram. The Aztecs started their expansion in 1325, so when Cortez arrived in 1519 they were a civilization 1519–1325 = 194 years old. Reporting this 194 years lapse before the year 3500, we find that the virtual Aztecs had been 'born' 194 years earlier, namely in 3694 B.C., which is thus the $b$-value of the virtual Aztec $b$-lognormal. Then we have to find the $b$-lognornal itself, i.e. its $\mu$ and $\sigma$. Its peak lies on the Greece-to-Spain exponential curve but, unfortunately, not exactly upon it since the system of two simultaneous equations (71) and (72) cannot be solved exactly for $\mu$ and $\sigma$. Thus, this approximated numerical solution, corresponding to the quadratic (77), is reflected in the diagram by positioning the virtual Aztec $b$-lognormal slightly above the Greece-to-Spain exponential. Finally, the true Aztec $b$-lognormal is just the same thing as the virtual Aztec $b$-lognormal except that its peak is shifted in time by an amount of (3500 + 1519) years = 5019 years into the future, so that its peak falls at 1519, when the Spaniards arrived.

One may let $\mu_{VA}$ disappear from the above two equations by multiplying them side-by-side and then finding the following new equation in $\sigma_{VA}$ only, that must thus be solved for $\sigma_{VA}$:

$$\text{Numerically\_known} = (p_{VA} - b_{VA})P_{VA} = \frac{e^{-\frac{\sigma_{VA}^2}{2}}}{\sqrt{2\pi}\sigma_{VA}}. \quad (75)$$

Unfortunately, it is not possible to solve this equation for $\sigma_{VA}$ exactly. The best we can do is to expand its right-hand side into a MacLaurin power series for $\sigma_{VA}$ (which is acceptable since we know that $0 < \sigma < 1$ for all $b$-lognormals representing life-spans), thus getting (the series is truncated at power 2 in $\sigma_{VA}$):

$$\text{Numerically\_known} = (p_{VA} - b_{VA})P_{VA} \approx \frac{1 - \frac{\sigma_{VA}^2}{2}}{\sqrt{2\pi}\sigma_{VA}}. \quad (76)$$

Solving this for $\sigma_{VA}$ leads to the quadratic in $\sigma_{VA}$

$$\sigma_{VA}^2 + 2\sqrt{2\pi}(p_{VA} - b_{VA})P_{VA}\sigma_{VA} - 2 = 0, \quad (77)$$

whose two roots are

$$\sigma_{VA} = -\sqrt{2\pi}(p_{VA} - b_{VA})P_{VA} \pm \sqrt{2}\sqrt{\pi(p_{VA} - b_{VA})^2 P_{VA}^2 + 1}. \quad (78)$$

Discarding the negative root, this leads to the only positive root for $\sigma_{VA}$

$$\sigma_{VA} = -\sqrt{2\pi}(p_{VA} - b_{VA})P_{VA} + \sqrt{2}\sqrt{\pi(p_{VA} - b_{VA})^2 P_{VA}^2 + 1} \quad (79)$$

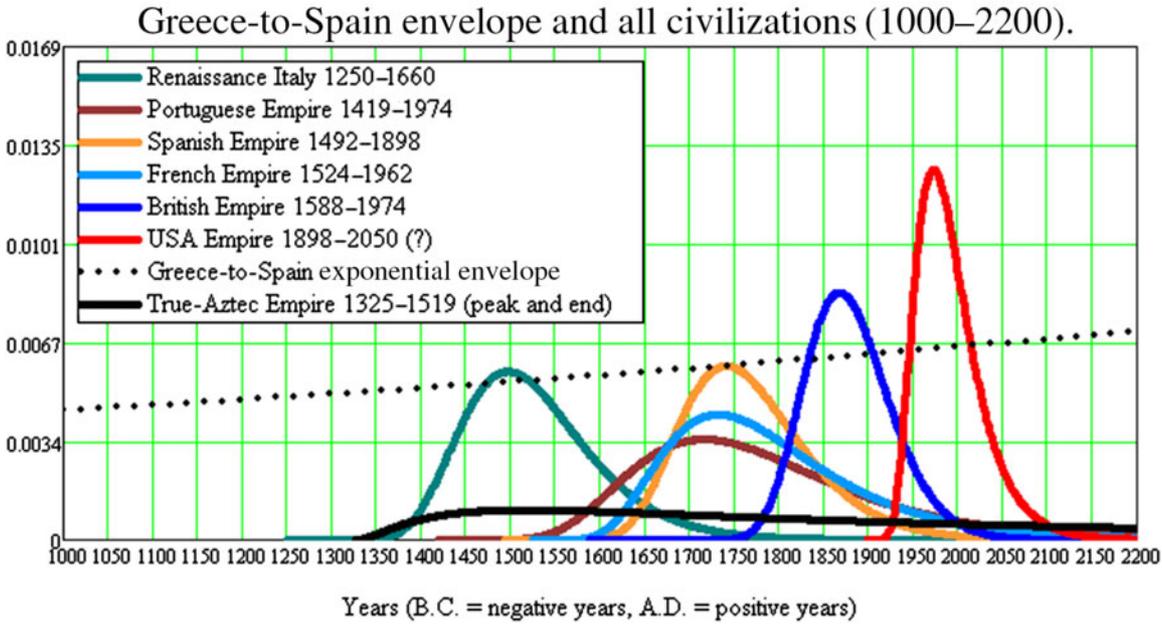

**Fig. 10.** Enlarged portion of Figure 9 limited to the years between 1000 and 2200.

and the problem is (approximately) solved, since $\mu_{VA}$ is then found from $\sigma_{VA}$ by solving the upper equation (74):

$$\mu_{VA} = \sigma_{VA}^2 + \ln(p_{VA} - b_{VA}). \tag{80}$$

Replacing the numerical values given by (73) and (74) into (79) and (80), the latter yield, respectively:

$$\begin{cases} \sigma_{VA} = 1.103, \\ \mu_{VA} = 6.484. \end{cases} \tag{81}$$

The numeric value for $\sigma_{VA}$ slightly higher than 1 is a bit surprising, and shows once again that we are working in a numerically approximated solution of the transcendental equation (75): a more accurate numeric solution of (75) would be needed. The same fact is revealed graphically in Figure 9, inasmuch as the virtual Aztec $b$-lognormal peak lies a little bit above the Greece-to-Spain exponential curve. In conclusion, the Virtual Aztec $b$-lognormal equation reads

Virtual_Aztec_$b$-lognormal$(t, \mu_{VA}, \sigma_{VA}, b_{VA})$

$$= \frac{1}{\sqrt{2\pi}\sigma_{VA}(t - b_{VA})} e^{-\frac{(\ln(t-b_{VA})-\mu_{VA})^2}{2\sigma_{VA}^2}}. \tag{82}$$

As for the true Aztec $b$-lognormal, that is just the same as the virtual Aztec $b$-lognormal except that it is shifted in time so as to start in 1325, when the Aztec expansion in central Mexico started. By construction, the peak of such true Aztec $b$-lognormal falls exactly in the year 1519, when the Spaniards arrived:

True_Aztec_$b$-lognormal$(t, \mu_{VA}, \sigma_{VA}, b_{TA})$

$$= \frac{1}{\sqrt{2\pi}\sigma_{VA}(t - b_{TA})} e^{-\frac{(\ln(t-b_{TA})-\mu_{VA})^2}{2\sigma_{VA}^2}}. \tag{83}$$

This true Aztec $b$-lognormal is also shown in Figure 11, just below all other $b$-lognormals, immediately revealing that the Aztecs were by far technologically inferior to all other European civilizations of the time. We shall now explore a little more in detail the last statement.

Consider Figures 10 and 11, which are enlarged portions of Figure 9 but limited to the years between 1000 and 2200 (Figure 10) and, even better, between the years 1300 and 1520 (Figure 11). Then one immediately infers that:

(1) Around 1497, the culturally leading country in the world was Renaissance Italy (green solid $b$-lognormal).
(2) In 1519, however, continental Spain (black dot-dot Greece-to-Spain exponential curve) had virtually reached the same cultural and technological level as the (already starting to decline) Renaissance Italy.
(3) In 1519, the Portuguese Empire (brown solid $b$-lognormal) was at its beginnings, since it had started in 1419.
(4) In 1519, the Spanish Empire (orange solid $b$-lognormal) was at its beginnings, since it had started in 1492.
(5) In 1519, the Aztec Empire (black solid $b$-lognormal) was at its top: ready to be crushed by the Spaniards, owing to the huge cultural and technological inferiority of the Aztecs (18.7%) to the Spaniards (assumed 100% in comparison).

## *b*-lognormal entropy as 'civilization amount'

*Introduction: invoking entropy and information theory*

We now take a more profound mathematical step ahead than just using $b$-lognormals: we resort to information theory, firstly put forward by Claude Shannon (1916–2001) in 1948.

In particular, we now need Shannon's notion of differential entropy $H$ of an assigned probability density $f_X(x)$, defined by the integral

$$H = -\int_{-\infty}^{\infty} f_X(x) \cdot \ln f_X(x)\, dx. \tag{84}$$

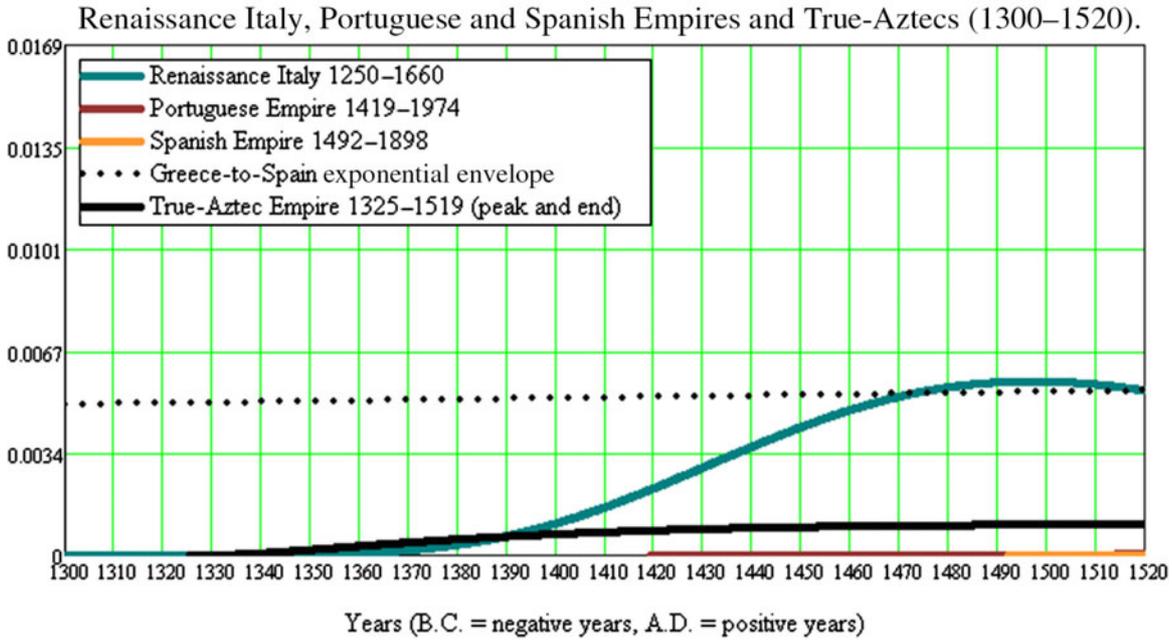

**Fig. 11.** Enlarged portion of Figure 10 limited to the years between 1300 and 1520. If we assume the technological level of the Spaniards to equal 100%, then the technological level of the Aztecs is only about 18%, i.e. the Aztec b-lognormal is about one-fifth of the Spaniard b-lognormal height in 1520. No wonder the Spaniards crushed the Aztecs, then. Yet, in the next section, we claim that Shannon's Information Theory provides an even better way to measure the cultural and technological gap between Aztec and Spaniards: this is what physicists have long been calling Entropy.

Essentially, this is a measure of 'how much peaked' a pdf is (small entropy) in contrast to a 'largely spread' pdf (high entropy), and so a pdf entropy is also called its 'uncertainty' (no relationship to Heisenberg's uncertainty principle).

In particular, we need the expression of differential entropy of the $b$-lognormal. Let us thus start by finding the differential entropy of the ordinary lognormal (i.e. starting at zero), that is:

$$\begin{aligned}H_{\text{lognormal}} &= -\int_0^\infty \frac{1}{\sqrt{2\pi}\sigma x}e^{-\frac{(\ln x-\mu)^2}{2\sigma^2}}\ln\left[\frac{1}{\sqrt{2\pi}\sigma x}e^{-\frac{(\ln x-\mu)^2}{2\sigma^2}}\right]dx\\ &= -\int_0^\infty \frac{1}{\sqrt{2\pi}\sigma x}e^{-\frac{(\ln x-\mu)^2}{2\sigma^2}}\left[-\ln\left(\sqrt{2\pi}\sigma\right)-\ln x-\frac{(\ln x-\mu)^2}{2\sigma^2}\right]dx\\ &= \ln\left(\sqrt{2\pi}\sigma\right)\int_0^\infty \frac{1}{\sqrt{2\pi}\sigma x}e^{-\frac{(\ln x-\mu)^2}{2\sigma^2}}dx + \int_0^\infty \ln x \frac{1}{\sqrt{2\pi}\sigma x}e^{-\frac{(\ln x-\mu)^2}{2\sigma^2}}dx\\ &\quad + \int_0^\infty \frac{(\ln x-\mu)^2}{2\sigma^2}\frac{1}{\sqrt{2\pi}\sigma x}e^{-\frac{(\ln x-\mu)^2}{2\sigma^2}}\end{aligned}$$

(85)

now the substitution $\ln x = z$ changes all three integrals into well-known integrals of normal distribution, equal to 1 (normalization condition), $\mu$ (mean value definition) and $\sigma^2$ (variance definition), respectively:

$$\begin{aligned}&= \ln\left(\sqrt{2\pi}\sigma\right)\int_{-\infty}^\infty \frac{1}{\sqrt{2\pi}\sigma}e^{-\frac{(z-\mu)^2}{2\sigma^2}}dz + \int_{-\infty}^\infty z\frac{1}{\sqrt{2\pi}\sigma}e^{-\frac{(z-\mu)^2}{2\sigma^2}}dz\\ &\quad + \int_{-\infty}^\infty \frac{(z-\mu)^2}{2\sigma^2}\cdot\frac{1}{\sqrt{2\pi}\sigma}e^{-\frac{(z-\mu)^2}{2\sigma^2}}dx\\ &= \ln\left(\sqrt{2\pi}\sigma\right)+\mu+\frac{1}{2\sigma^2}\sigma^2 = \ln\left(\sqrt{2\pi}\sigma\right)+\mu+\frac{1}{2}.\end{aligned}$$

As for the differential entropy of the $b$-lognormal, it is just the same as (85), since the $b$-lognormal simply is a lognormal shifted to a new origin $b$ along the $x$-axis, and so all infinite-support integrals in (85) remain unchanged, and the proof is just the same as (85). Thus, in conclusion, the differential entropy of both the ordinary lognormal and the $b$-lognormal is given by

$$H_{\text{lognormal}} = H_{b\text{-lognormal}} = \ln\left(\sqrt{2\pi}\sigma\right)+\mu+\frac{1}{2}. \quad (86)$$

Notice also that (86) yields the $b$-lognormal differential entropy in nats, i.e. in natural logarithms. If one wants to express (86) in bits, then one must divide (86) by $\ln 2 = 0.693\ldots$ In other words, one has

$$\begin{aligned}H_{\text{lognormal\_in\_bits}} &= H_{b\text{-lognormal\_in\_bits}} = \frac{1}{\ln 2}\left[\ln\left(\sqrt{2\pi}\sigma\right)+\mu+\frac{1}{2}\right]\\ &\approx 1.443\ldots\left[\ln\left(\sqrt{2\pi}\sigma\right)+\mu+\frac{1}{2}\right].\end{aligned}$$

(87)

The proof of (87) from (86) simply follows from the well-known change-of-base formula for the logarithms

$$\log_2 N = \frac{\ln N}{\ln 2} \approx \frac{\ln N}{0.693} \approx 1.443 \ln N \quad (88)$$

applied to definition (84) of differential entropy with $\ln(\ldots)$ replaced by $\log_2(\ldots)$.

*Exponential curve in time determined by two points only*

Let us now rewrite the exponential curve in time:

$$E(t) = Ae^{Bt}. \quad (89)$$

Now suppose that the exponential curve (89) is passing through two (and only two) assigned points of coordinates ($p_1$, $P_1$) and ($p_2$, $P_2$), respectively. In other words, we assume that the two simultaneous equations hold

$$\begin{cases} P_1 = Ae^{Bp_1} \\ P_2 = Ae^{Bp_2}. \end{cases} \quad (90)$$

This system (90) of two equations may be solved with respect to the two constants $A$ and $B$ in a few simple steps that we omit here (but they are found in the Appendix 30. A of Maccone (2012), see equations %02 through %07). The result is:

$$\begin{cases} A = \dfrac{P_1^{\frac{p_2}{p_2-p_1}}}{P_2^{\frac{p_1}{p_2-p_1}}} \\ B = \dfrac{\ln\left(\frac{P_2}{P_1}\right)}{p_2 - p_1}. \end{cases} \quad (91)$$

Thus, the two equations (91) completely solve the problem of finding the exponential curve in time (89) passing through the two assigned points of coordinates ($p_1$, $P_1$) and ($p_2$, $P_2$).

*Assuming that the exponential curve in time is the GBM mean value curve*

We are now ready to take the usual step ahead in our stochastic representation of Darwinian theory (Evolution, as described above) and of historical progress (Mathematical History, as described in 'Extrapolating history into the past: Aztecs' section) assuming that the exponential curve in time (89) is the same thing as the exponential *mean value* of GBM in time, given by (15) (and (11.1) in Maccone (2012)), that is

$$\langle N(t)\rangle = N_0 e^{\mu t} = N_0 e^{\mu_{GBM} t}. \quad (92)$$

Evolution and human progress now cease to be deterministic chains of events and rather become a stochastic (random) sequence of events, as indeed it is in reality, with all their ups and downs, although constrained to have a deterministic exponential overall *average* increase. This 'mathematical representation of Darwinian Evolution and Historic Human Progress as Geometric Brownian Motion' is thus the #1 message put forward by this paper and by the Maccone (2012) book, while the use of entropy is the #2 message.

In equation (92) we have set

$$\mu = \mu_{GBM} \quad (93)$$

to remind that the drift parameter $\mu$ is now the drift parameter of the GBM, denoted by $\mu_{GBM}$.

Having so said, let us now check (89) against (92). One then obtains

$$\begin{cases} A = N_0, \\ B = \mu_{GBM}. \end{cases} \quad (94)$$

On the other hand, it will be remembered from (26) (and (11.12) of Maccone (2012)) that the relationship between $N_0$ and $\sigma_{GBM}^2$ is

$$N_0 = e^{\frac{\sigma_{GBM}^2}{2}}. \quad (95)$$

Thus, upon inserting (95) into the upper equation (94), the latter may be rewritten

$$\begin{cases} A = e^{\frac{\sigma_{GBM}^2}{2}} \\ B = \mu_{GBM}. \end{cases} \quad (96)$$

Solving then (96) for both $\mu_{GBM}$ and $\sigma_{GBM}^2$ one obtains:

$$\begin{cases} \mu_{GBM} = B, \\ \sigma_{GBM}^2 = 2\ln A. \end{cases} \quad (97)$$

These two equations may be rewritten in terms of the two points of coordinates ($p_1$, $P_1$) and ($p_2$, $P_2$) by inserting (91) into (97). With a little rearranging, the result is

$$\begin{cases} \mu_{GBM} = \dfrac{\ln\left(\frac{P_2}{P_1}\right)}{p_2 - p_1} \\ \sigma_{GBM}^2 = \dfrac{2\ln\left(\frac{P_1^{p_2}}{P_2^{p_1}}\right)}{p_2 - p_1}. \end{cases} \quad (98)$$

In conclusion, we have:
(1) Determined the exponential curve in time passing through the two assigned points of coordinates ($p_1$, $P_1$) and ($p_2$, $P_2$). These two points will later be identified as the two peaks of the initial b-lognormal ($p_1$, $P_1$) and final b-lognormal ($p_2$, $P_2$), respectively.
(2) Assumed that the above exponential curve in time is the mean value of a certain GBM. This allows for the representation of Darwinian evolution and later human history as a GBM, with all its ups and downs, but constrained by definition to have an exponential increase in the 'level of evolution', as 3.5 billion years of evolution and progress on Earth plainly show.
(3) Found the two equations (98) fully expressing this GBM in terms of the two initial and final peaks only. This is in full agreement with the statistical Drake equation static case, as described in 'GBM as the key to stochastic evolution of all kinds' and 'Darwinian Evolution re-defined as a GBM in the number of living species' section (and in Chapters 1, 3, 6, 7, 8 and 11 of the author's book, Maccone (2012)).

The way is thus paved to outline a full mathematical (statistical) theory of Darwinian evolution and human history, later to be extended to alien civilizations after the first 'Contact', even if these ETs will be 'post-biological' (namely based on artificial intelligence, rather than on 'flesh').

*The 'No-Evolution' stationary stochastic process*

This short section is devoted to the special case when the two peak ordinates, $P_1$ and $P_2$, are equal to each other:

$$P_2 = P_1. \quad (99)$$

Then, (99) and (91), with a little rearrangement, yield

$$\begin{cases} A = P_1 = P_2, \\ B = 0. \end{cases} \quad (100)$$

This means that the 'former exponential' is now a straight line parallel to the time axis and located at a certain positive ordinate $A = P_1 = P_2$ above it. In other words, the former GBM stochastic process has now become a stationary stochastic process, showing that 'No-Evolution' occurs between the two times $p_1$ and $p_2$: for millions or billions of years, living beings are born, reproduce and die generation after generation, with no evolution at all.

Also, the drift parameter $\mu_{GBM}$ vanishes, and the $\sigma^2_{GBM}$ is a constant in time equal to twice the natural log of the constant peak ordinate $(P_1, P_2)$, as one immediately may see by inserting (99) into (98):

$$\begin{cases} \mu_{GBM} = 0, \\ \sigma^2_{GBM} = 2\ln(P_1) = 2\ln(P_2). \end{cases} \quad (101)$$

*Entropy of the 'Running b-lognormal' peaked at the GBM exponential mean*

Much more interesting than the 'No-Evolution' case just considered is of course the truly exponential evolution case. It is summarized at-a-glance in Figure 4 that intuitively shows the basic mechanism that we propose in this paper, in order to unify the notions of Darwinian evolution, human historical progress and SETI. However, all this is just qualitative.

If we want to go quantitative, the only way is to resort to Shannon's information theory and then resort to the notion of *b*-lognormal differential entropy described in Section 'Introduction: invoking entropy and information theory', as we now describe in detail.

To start, let us call 'running *b*-lognormal' the generic *b*-lognormal peaked at the generic instant $t = p$.

It will be remembered that the key equations describing the running *b*-lognormal are the two equations (31) that we reproduce here conveniently re-written in this section's notation (the subscript '*RbL*' means 'running *b*-lognormal'):

$$\begin{cases} \mu_{RbL} = \dfrac{1}{4\pi A^2} - pB, \\ \sigma_{RbL} = \dfrac{1}{\sqrt{2\pi} A}. \end{cases} \quad (102)$$

Therefore, the differential entropy of the running *b*-lognormal in bits is found by inserting (102) into (87), with the immediate result

$$H_{\text{running\_b-lognormal\_in\_bits}} = \frac{\ln(\sqrt{2\pi}\sigma_{RbL}) + \mu_{RbL} + \frac{1}{2}}{\ln 2}$$
$$= \frac{-\ln(A) + \frac{1}{4\pi A^2} - pB + \frac{1}{2}}{\ln 2}. \quad (103)$$

This equation expresses the running *b*-lognormal's differential entropy in terms of the two constants $A$ and $B$ of the exponential curve (89) and of the abscissa in time (i.e. $p$) at which the running *b*-lognormal's peak is positioned. Hence, in reality, the only 'free parameter' in (103) is indeed this 'free' *b*-lognormal's peak abscissa $p$. Let us write this fact neatly as follows:

$$H_{\text{running\_b-lognormal\_in\_bits}}(p) = \frac{-\ln(A) + \frac{1}{4\pi A^2} - pB + \frac{1}{2}}{\ln 2}. \quad (104)$$

We may now rewrite the last equation in terms of the coordinates of the two points $(p_1, P_1)$ and $(p_2, P_2)$ by inserting (91) into (104). The result of this straight substitution is:

$$H_{\text{running\_b-lognormal\_in\_bits}}(p)$$
$$= \frac{1}{\ln 2}\left[-\ln\left(\frac{P_1^{\frac{p_2}{p_2-p_1}}}{P_2^{\frac{p_1}{p_2-p_1}}}\right) + \frac{1}{4\pi\left(\frac{P_1^{\frac{p_2}{p_2-p_1}}}{P_2^{\frac{p_1}{p_2-p_1}}}\right)^2} - p\frac{\ln\left(\frac{P_2}{P_1}\right)}{p_2-p_1} + \frac{1}{2}\right]. \quad (105)$$

This formula may only slightly be simplified as follows:

$$H_{\text{running\_b-lognormal\_in\_bits}}(p)$$
$$= \frac{1}{\ln 2}\left[-\frac{\ln\left(\frac{P_1^{p_2}}{P_2^{p_1}}\right)}{p_2-p_1} + \frac{P_2^{\frac{2p_1}{p_2-p_1}}}{4\pi P_1^{\frac{2p_2}{p_2-p_1}}} - p\frac{\ln\left(\frac{P_2}{P_1}\right)}{p_2-p_1} + \frac{1}{2}\right]. \quad (106)$$

This is the differential entropy in bits of the running *b*-lognormal peaked at *p*. It may more conveniently be rewritten as

$$H_{\text{running\_b-lognormal\_in\_bits}}(p) = -p\frac{\ln\left(\frac{P_2}{P_1}\right)}{(p_2-p_1)\ln 2}$$
$$+ \text{Part\_not\_depending\_on\_}p. \quad (107)$$

We will use (107) in a moment!

*Decreasing entropy for an exponentially increasing evolution: progress!*

We now reach the conclusion of this paper, as well as the Epilogue of the book on 'Mathematical SETI'.

Young students at university courses on thermodynamics are still made to learn that 'entropy always increases', meaning that the second law of thermodynamics rules so. However, thermodynamics was born in 1700–1800 days upon the discovery of the laws of gases, and quite a few scientists have later (1900–2000 days) come to realize that the sentence 'entropy always increases' may hardly apply to the evolution of intelligent living species as it occurred on Earth over the last 3.5 billion years. This author belongs to the latter category of scientists, and is proud to claim that his theory, outlined in this paper as well as in Chapters 6, 7, 8 and 30 of his book, neatly shows that entropy decreases (rather than always increasing) when it comes to describe the evolution of life up to the current time, i.e. when it comes to describe progress!

Our proof is as follows.

Consider (87), i.e. the entropy of the running *b*-lognormal.

If we consider the entropy change between the initial *b*-lognormal peaked at $(p_1, P_1)$ and the final one peaked at $(p_2, P_2)$, this entropy change is, by definition, given by

$$\Delta_{H\_in\_bits} = H_{\text{running\_b-lognormal\_in\_bits}}(p_2)$$
$$- H_{\text{running\_b-lognormal\_in\_bits}}(p_1). \quad (108)$$

When rewritten in terms of (107), this entropy change is simpler than (107) since the two 'Parts_not_depending_on_$p$' cancel against each other, and the result is just:

$$\Delta_{H\_in\_bits} = H_{\text{running\_}b\text{-lognormal\_in\_bits}}(p_2) \\ - H_{\text{running\_}b\text{-lognormal\_in\_bits}}(p_1) \\ = -p_2 \frac{\ln\left(\frac{P_2}{P_1}\right)}{(p_2-p_1)\ln 2} + p_1 \frac{\ln\left(\frac{P_2}{P_1}\right)}{(p_2-p_1)\ln 2}. \quad (109)$$

However, the last equation may be further simplified as follows:

$$\Delta_{H\_in\_bits} = -p_2 \frac{\ln\left(\frac{P_2}{P_1}\right)}{(p_2-p_1)\ln 2} + p_1 \frac{\ln\left(\frac{P_2}{P_1}\right)}{(p_2-p_1)\ln 2} \\ = -(p_2-p_1)\frac{\ln\left(\frac{P_2}{P_1}\right)}{(p_2-p_1)\ln 2} = -\frac{\ln\left(\frac{P_2}{P_1}\right)}{\ln 2} = \frac{\ln\left(\frac{P_1}{P_2}\right)}{\ln 2}. \quad (110)$$

Thus, we have reached a result of adamantine beauty: the entropy change, when passing from a lower civilization to a higher civilization, is simply given by the log of the ratio between the lower civilization peak and the higher civilization peak (apart from the factor ln 2 at the denominator, necessary to measure the entropy in bits):

$$\Delta_{H\_in\_bits} = \frac{\ln\left(\frac{P_1}{P_2}\right)}{\ln 2}. \quad (111)$$

Since in the actual unfolding of evolution, we always had $P_1 < P_2$, the entropy change (111) in passing from the lower civilization to the higher civilization is the log of a number smaller than 1, i.e. it is a negative number! Thus, entropy decreases in passing from a lower civilization to a higher civilization, and that proves that progress amounts to a decrease in entropy.

One more important result derived from (107) is found upon rewriting it explicitly as a function of $p$ as follows:

$$H_{\text{running\_}b\text{-lognormal\_in\_bits}}(p) = \frac{-\ln(A) + \frac{1}{4\pi A^2} - p\,B + \frac{1}{2}}{\ln 2} \\ = \frac{-B}{\ln 2}p + \text{part\_not\_depending\_on\_}p. \quad (112)$$

Then, (108) and (112) yield immediately

$$\Delta_{H\_in\_bits} = -\frac{B}{\ln 2}(p_2 - p_1). \quad (113)$$

This also is a new result of adamantine beauty, since it yields the entropy change in bits by virtue of the difference in time between the two peak abscissas ($p_2 - p_1$) and the constant $B$ expressing the exponential increase (i.e. the mean exponential drift in the GBM) of (89). Just to distinguish the two adamantine results (111) and (113) from each other, we might call (111) the 'entropy change in evolution by virtue of the peak ordinates only', and (113) the 'entropy change in evolution by virtue of the peak abscissas and the average GBM exponential drift, $B$, only'.

*Six examples: entropy changes in Darwinian Evolution, Human History between ancient Greece and now, and Aztecs and Incas versus Spaniards*

Appendix 30.A of Maccone (2012), not only provides full mathematical proofs of all the results we have derived so far in this paper: but also offers six examples showing how entropy actually decreased in six different cases of past life on Earth.

These six cases are, respectively:

(1) Darwinian Evolution on Earth over the last 3.5 billion years. This is shown to correspond to an entropy decrease of 25.57 bits per each living being, if today's number of living species is assumed to be 50 million. Were there more than 50 million species living on Earth right now, our equations (111) and (113) would yield the corresponding entropy decrease accordingly. As for the proofs, please refer to equations (%i45) through (%o52) of Appendix 30.A of Maccone (2012).

(2) Entropy changes in human history from ancient Greece to the end of the British Empire, i.e. to nearly nowadays. According to the $b$-lognormal theory that we described in Section 'Extrapolating history into the past: Aztecs', the year 434 B.C. (i.e. $p_1 = -434 \cdot \text{years}$) corresponds to the peak of the age of Pericles in Athens, while the year 1868 (i.e. $p_2 = 1868 \cdot \text{years}$) corresponds to the peak of the Victorian age in Britain (Maxwell equations mastering electromagnetism published just 4 years earlier, in 1864). The entropy change between these two peaks is computed in equations (%i53) through (%o59) of Maccone (2012), and amounts to an entropy decrease of 1.76 bits per individual.

(3) Entropy changes in human history from ancient Greece to the end of the American Empire, assumed to yield to China about the year 2050. Again, according to the $b$-lognormal theory that we described in Section 'Extrapolating history into the past: Aztecs', the year 434 B.C. (i.e. $p_1 = -434 \cdot \text{years}$) corresponds to the peak of the age of Pericles in Athens, while the year 1973 (i.e. $p_2 = 1973 \cdot \text{years}$) corresponds to the peak of the American Empire (Americans had just landed on the Moon on July 20th, 1969, and stopped landing on 14 December 1972). Equations (%i60) through (%o66) of Maccone (2012) show that the corresponding entropy decrease between the two peaks amounts to 2.38 bits per individual, thus a higher number than for the Greece-to-Britain GBM exponential mean, as is intuitively obvious.

(4) Computing the entropy difference between the Aztecs and the Spaniards when they came suddenly in touch (with no previous contact) in 1519, when Cortez landed in Vera Cruz and started invading the Aztec Empire. This example is particularly important for SETI inasmuch as it 'might resemble' what could happen in case humanity came physically in touch with an Alien Civilization all of a sudden (as shown in the movie 'Independence Day'). Well, in section ''Virtual-Aztecs' method to find the 'True-Aztecs' $b$-lognormal' we found a way to compute 'how backward the Aztecs were' by exploiting the fact that they were just one the verge of discovering the use of wheels. Wheels were used by Aztec children in their toys, but not by adults in warfare, and this is just what had already happened in Mesopotamia, Northern Caucasus and Central Europe about 3500 years before Christ. Thus, we set $p_1 = -3500 \cdot$ years to describe the Aztec technological backwardness with respect to the Spaniards, and of course we set $p_2 = 1521 \cdot$ years to mean that Cortez completed the conquest of the Aztec Empire in 1521. Again, we assumed the Greece-to-Britain exponential to be the right one (since it refers to the actually occurred facts, rather than future facts also, as in the Greece-to-USA case). In conclusion, (111) and (113) then yield a entropy difference of 3.84 bits per individual between the Aztecs and the Spaniards. This is higher than the 1.76 bits difference between the Victorian Britons and the Pericles Greeks, of course: nearly twice as much, owing to the huge 5000-years difference in evolution between the Aztecs and the Spaniards, versus the just about 2300 years in evolution between Pericles and the Victorian age.

Let us also compute the entropy difference between the Incas and the Spaniards when they came suddenly in touch (with no previous contact) in 1532, when Pizarro landed in Tumbes and started invading the Incas Empire. Again, the wheel was unknown to the Incas, and so we assumed $p_1 = -3500 \cdot$ years. Naturally, we now assumed $p_2 = 1532 \cdot$ years, and the result given by (111) and (113) is an entropy difference of 3.85 bits per individual. Just very slightly higher than for the Aztec–Spaniard case, meaning that . . .

(5) The Aztecs were very slightly more technologically advanced than the Incas when they both were subdued by the Spaniards. In fact, if you assume that the Aztec Empire had its start around the year 850 A.D. (roughly at the peak of the Maya Empire, so that, in some sense, the Aztec 'inherited' part of the Maya civilization), and then you assume that the Incas Empire was founded around 1250 (when the Incas reached Cuzco), then, assuming again the Greece-to-Britain exponential as the true exponential of technological development, (111) and (113) yield an entropy difference of 0.3 bits per individual in favour of the Aztecs (more technologically advanced) over the Incas. A subtle quantitative statement that may be the current historical knowledge about both people is possibly unable to ponder over.

Hence, with these final six entropy measurements, we hope to have been the first author to be able to give a quantitative description of both Darwinian evolution and human history, based on our new discoveries about the mathematical properties of the finite and infinite $b$-lognormals.

This we did to be able to quantitatively estimate how much an alien civilization might be more advanced than us.

b-*lognormals of alien civilizations*

So much about the past, but what about the future?

What are the $b$-lognormals of ET civilizations?

Nobody knows, of course, and nobody will until the SETI scientists will detect the first signs of an ET civilization.

A good book to read, however, is 'Interstellar Migration and the Human Experience', by Finney, B. R. and Jones, E. M. (1986). Also, science fiction fans might take pleasure in casting the Star Trek timeline into the mathematical language of our $b$-lognormals. Interesting is also 'The Star Trek Chronology' by Okuda & Okuda (1996), but . . . no mathematics is there. We need the mathematics that we will develop in our next paper by extrapolating exponentials and entropy of the human past into the future, with reference to the Fermi Paradox.

**Conclusion: summary of technical concepts described**

As a conclusion to this paper, we would like to summarize the new technical concepts we had to introduce here.

(1) The Drake equation, describing 10 billion years of evolution in this Galaxy (stars to humans), was transformed from the simple product of seven factors into the product of any number of random variables. This statistical Drake equation is more 'serious' scientifically, and leads to the conclusion that, if the number of input random variables is increased more and more, then the pdf of the number of civilizations in the Galaxy must be lognormal.

(2) Darwinian evolution on Earth was re-defined mathematically as a stochastic process (i.e. a random function of time) yielding the number of living species on Earth over the last 3.5 billion years. This definition allows for sudden lows in the number of living species (mass extinctions) but, apart from those, the overall mean behaviour of the number of species of Earth in time must be increasing exponentially. Today, some 50 million species are supposed to live on Earth, while 3.5 billion years ago there was just one (RNA?), thus fixing the exponential mean curve perfectly.

(3) Geometric Brownian Motion (GBM) is exactly the right stochastic process fulfilling all the above requirements representing the evolution of life on Earth (and elsewhere in the universe, like life on extrasolar planets). However, GBMs were so far studied only in Financial Mathematics (Black–Scholes models, leading to the Nobel Prize in Economy assigned in 1997 to Scholes and Merton for exploiting GBMs), and so it is high time for evolutionary scientists and astrobiologists to realize the key merits of GBMs, primarily their mean value increasing in time exponentially.

(4) We took one more step ahead by introducing lognormal distributions ($b$-lognormals) starting at any positive time $b > 0$ rather than just at zero, as ordinary lognormals do. Then, these $b$-lognormals were 'matched' to GBMs by forcing all their peaks to stay just on the GBM exponential mean value curve. This leads to a one-parameter family of $b$-lognormals (the parameter is the peak abscissa $p$) representing a new living species that appeared on Earth exactly at time $b$.

(5) Cladistics, the science of Evolution Phylogenetic Trees, then is reduced to a simple game of $b$-lognormals departing from the main exponential curve of evolution (i.e. the GBM mean value) and then either increasing, or decreasing, or keeping constants in time but in a stochastic fashion (this is our 'NoEv' new pdf, that is, not a lognormal any more). In other words, we have accounted for prospering species, or extinct species (decreasing exponential arches that, sooner or later, reach a numeric value above zero but less than one, meaning extinction), or even just 'stationary' species (like insects, for instance, that keep being the same as they were about 400 million years ago).

(6) The lifetime of any living being may also be represented by a made-finite $b$-lognormal. In fact, every living being is born at time $b$, reaches puberty at time $a$ (adolescence, i.e. the abscissa of the $b$-lognormal increasing inflexion point), then goes to the peak of his living capabilities (abscissa $p$ of the $b$-lognormal's peak) and starts declining. He/she then reaches the non-return decline point (abscissa $s$ of the 'senility' point, the decreasing inflection point abscissa) and dies at time $d$ when the straight line tangent to the $b$-lognormal at senility intercepts the time axis.

(7) The 'golden ratio' was long hailed by artists, architects and men of literature as 'symbol of visual perfection'. Well, surprisingly enough, this author discovered a class of $b$-lognormals strongly related to the golden ratio (golden $b$-lognormals). The future will show whether this discovery is just an iceberg's tip, leading to many more discoveries related to the golden ratio's already quite rich literature.

(8) However, $b$-lognormals cannot be used to describe the lifetime of a living being only. They may be used to describe the lifetime of societies also. There is a profound theorem behind all this, stating that (in easy terms) 'just as the sum of two independent Gaussian distributions is one more Gaussian, similarly the product of two independent lognormal distributions is one more lognormal'. We thus could use $b$-lognormals to study the historic course of human civilizations on Earth (i.e. the 'f sub i' factor in the Drake equation).

(9) The history of Ancient Greece, Ancient Rome, Renaissance Italy, and then Portugal, Spain, France, Britain and the USA Empires were then cast into the language of $b$-lognormals. This was possible since the author discovered two equations ('History Formulae') that allow the computation of the $b$-lognormal's μ and σ given the birth time $b$, the death time $d$ and the intermediate value of 'senility $s$' (incipient decline), where the (infinite) $b$-lognormal hinges with the straight line going to death, thus making the infinite $b$-lognormal a finite one, as all lives are. Also, the $b$-lognormal of the Aztec civilization was computed as an essay in mathematical history.

(10) However, the most important result achieved by this author is undoubtedly his study of Entropy as 'Civilization Amount'. In fact, each $b$-lognormal has a precise entropy (or 'uncertainty') value in the sense of Shannon's information theory and so, for instance, it is possible to assign entropy values to all historic civilizations previously represented by virtue of $b$-lognormals: Aztecs, Greece, Rome, Renaissance Italy, Portugal, Spain, France, Britain and the USA. This explains by entropy values, rather than by just words, 'how much' a civilization was more or less 'organized' (i.e. 'advanced') than another. For instance, the entropy difference between Aztecs and Spaniards, when they clashed in 1519-20-21, turned out to equal about 3.85 bits per individual, while the entropy difference between the first living being of 3.5 billion years ago (RNA?) and today's humans turned out to equal 27.57 bits for each living being, thus providing a direct numeric measure of different evolving species or civilizations. This author plans to investigate these new results more in depth in forthcoming papers.

In conclusion, this author thinks he could make true progress by casting Evolution, Human History and SETI into his unified statistical framework made up of $b$-lognormals and GBMs. Quite simply, this was possible since he avoided sterile philosophical debates like 'What is Life?' and replaced them by the theme 'WHEN did Life occur?'.